\documentclass[fleqn,usenatbib,12pt]{mnras}

\usepackage[T1]{fontenc}
\usepackage{ae,aecompl}

\usepackage{graphicx}	
\usepackage{subfigure}	
\usepackage{amsmath}	
\usepackage{amssymb}	
\usepackage{bm}
\usepackage{ulem}

\newcommand{\zoh}{12+\log \mathrm{(O/H)}}
\newcommand{\irx}{L_{\rm IR}/L_{\rm UV}}
\newcommand{\oii}{[O\,{\small II}]$\lambda$3727}
\newcommand{\oiii}{[O\,{\small III}]$\lambda$5007}
\newcommand{\nii}{[N\,{\small II}]$\lambda$6584}

\title[A Universal Relation of Dust Obscuration]{A Universal Relation of Dust Obscuration Across Cosmic Time}
\author[J. Qin et al.]{
Jianbo Qin,$^{1,2}$ \thanks{E-mail: jbqin@pmo.ac.cn (JQ); xzzheng@pmo.ac.cn (XZZ)}
Xian~Zhong~Zheng,$^{1}$
Stijn Wuyts,$^{3}$
Zhizheng~Pan$^{1}$
and Jian~Ren$^{1,2}$
\\
$^{1}$Purple Mountain Observatory, Chinese Academy of Sciences, 8 Yuanhua Road, Nanjing 210034, China\\
$^{2}$School of Astronomy and Space Sciences, University of Science and Technology of China, Hefei 230026, China \\
$^{3}$ Department of Physics, University of Bath, Claverton Down, Bath BA2 7AY, UK}

\date{Accepted 2019 March 8. Received 2019 March 8; in original form 2018 November 22}
\pubyear{2019}
\begin{document}
\label{firstpage}
\pagerange{\pageref{firstpage}--\pageref{lastpage}}
\maketitle

\begin{abstract}

We investigate dust obscuration as parameterised by the infrared excess IRX$\equiv$$\irx$ in relation to global galaxy properties, using a sample of $\sim$32\,000 local star-forming galaxies (SFGs) selected from SDSS, GALEX and WISE. We show that IRX generally correlates with stellar mass ($M_\ast$), star formation rate (SFR), gas-phase metallicity ($Z$), infrared luminosity ($L_{\rm IR}$) and the half-light radius ($R_{\rm e}$). A weak correlation of IRX with axial ratio (b/a) is driven by the inclination and thus seen as a projection effect. 

By examining the tightness and the scatter of these correlations, we find that SFGs obey an empirical relation of the form $IRX$=$10^\alpha\,(L_{\rm IR})^{\beta}\,R_{\rm e}^{-\gamma}\,(b/a)^{-\delta}$ where the power-law indices all increase with metallicity. The best-fitting relation yields a scatter of $\sim$0.17\,dex and no dependence on stellar mass. Moreover, this empirical relation also holds for distant SFGs out to $z=3$ in a population-averaged sense, suggesting it to be universal over cosmic time. Our findings reveal that IRX approximately increases with $L_{\rm IR}/R_{\rm e}^{[1.3 - 1.5]}$ instead of $L_{\rm IR}/R_{\rm e}^{2}$ (i.e., surface density). We speculate this may be due to differences in the spatial extent of stars versus star formation and/or complex star-dust geometries. We conclude that not stellar mass but IR luminosity, metallicity and galaxy size are the key parameters jointly determining dust obscuration in SFGs.
\end{abstract}

\begin{keywords}
galaxies: obscuration -- galaxies: interstellar medium -- galaxies: dust -- galaxies: high-redshift
\end{keywords}

\large
\section{INTRODUCTION} \label{sec:sec1}

Dust plays a critical role in many aspects of galaxy evolution. It is produced during late stages of stellar evolution and is closely linked to the gas-phase metals in a galaxy. Dust absorbs the ultraviolet (UV) radiation and thermally re-emits into the infrared (IR). This effect, namely dust obscuration, influences many observables of galaxies. For star-forming galaxies (SFGs), dust obscuration can be parameterised by the ratio of IR to UV luminosity (i.e., the IR excess IRX$\equiv$$\irx$) \citep{Meurer1999,Heckman1998,Martin2005}. The IR excess is generally correlated with the Balmer decrement \citep{GB2010,Xiao2012,Koyama2018} and both of them often relate to the dust column density, or gas column density at a given metallicity in galaxies \citep{Bohlin1978,Guver2009,Zhu2017}. Identifying the physical quantities governing dust obscuration is key to understanding the dust evolution in relation to the metal enrichment of the interstellar medium (ISM) in galaxies and feeds into an empirical picture of galaxy evolution more broadly. 

A relationship of IRX with stellar mass among SFGs has been established out to $z$$\sim$4 \citep{Heinis2014,Whitaker2014,Pannella2015,AM2016},  revealing that low-mass SFGs with $M_\ast$$<$$10^{10.5}$\,$M_\odot$ evolve little in IRX  while high-mass SFGs significantly increase in IRX out to $z$=3. Similar results were obtained using the Balmer decrement as a proxy of dust obscuration \citep{Kashino2013,Price2014,Zahid2014b}.

In contrast, the typical star formation rate (SFR) of the low-mass SFGs decreases by a factor of $\sim$20 since $z$$\sim$3 \citep{Karim2011,Whitaker2014,Schreiber2015}. Their sizes and metallicities also undergo significant evolution over the same redshift range \citep{VDW2014,Maiolino2008,Zahid2014a}. Assuming a homogeneous mixture geometry in combination with the Kennicutt-Schmidt (K-S) Law, \citet{Wuyts2011} contrasted IRX values predicted based on $\Sigma_{\rm SFR}$ and metallicity to the observations, finding a good match at $z \sim 0$, but systematic overpredictions of the degree of obscuration compared to the observed IRX at higher redshifts. This discrepancy is also confirmed by \citet{Genzel2013} who found typical optical depths in high-$z$ SFGs to be $\sim$5 times lower than that of their local counterparts with a similar gas surface density. These puzzling results demonstrate that the fundamental quantities setting the level of dust obscuration in SFGs remain to be understood.

Dust obscuration relies not only on the dust surface density but also on the geometry of dust versus young stars in a galaxy \citep[see, e.g.,][]{Calzetti1994,CF2000,Li2019}. Investigating the scaling relations between dust obscuration and galaxy properties sheds light into how the intrinsic radiation is obscured. The correlation between dust obscuration and metallicity suggests that galaxies of higher metallicity usually have a higher dust-to-gas ratio, and then higher dust obscuration  \citep{Heckman1998,Johnson2007}. Dust obscuration also increases with SFR or SFR surface density \citep{Wang1996, Martin2005, GB2010, Xiao2012}, as well as galaxy stellar mass  \citep{GB2010,Reddy2010,Price2014,Whitaker2014,Zahid2017}. In addition, the observed dust obscuration is weakly correlated with the inclination of disc SFGs due to enhanced projected dust columns \citep{Wild2011,Chevallard2013,Leslie2018}.

 However, it is still under debate what parameters of galaxies are vital to determining dust obscuration.  \citet{GB2010} examined the dependence of dust obscuration on stellar mass, SFR and metallicity using a sample of local SFGs and argued that galaxy stellar mass is the most fundamental parameter.  On the other hand, \citet{Xiao2012} pointed out that a conjunction of multiple parameters  does a better job than a single parameter in interpreting dust obscuration as functions of H$\alpha$ luminosity (or SFR) surface density, metallicity and axial ratio for nearby SFGs. Moreover, a full description of the causes for dust obscuration should be able to account for the observed obscuration levels in both local and distant SFGs. Such efforts are still lacking.

In this work, we aim to identify the fundamental parameters determining dust obscuration in galaxies by addressing the relationships between dust obscuration and other galaxy properties using samples of local SFGs, as well as those out to high $z$.

In Section~\ref{sec:sec2}, we present the data, the local sample of SFGs and estimate of galaxy parameters. Section~\ref{sec:sec3} shows the correlations between dust obscuration and different galaxy parameters. We derive a new empirical relation of IRX in Section~\ref{sec:sec4} and apply this relation to high-$z$ galaxies in Section~\ref{sec:sec5}.  We discuss our results in Section~\ref{sec:sec6} and give a summary in Section~\ref{sec:sec7}. A $\Lambda$CDM cosmology with $H_0$=70\,km$^{-1}$\,Mpc$^{-1}$, $\Omega _{\rm \Lambda} = 0.7$ and $\Omega _{\rm m} = 0.3$ and a \citet{Chabrier2003} Initial Mass Function (IMF) are adopted unless mentioned otherwise.

\begin{figure}
\includegraphics[width=0.48\textwidth]{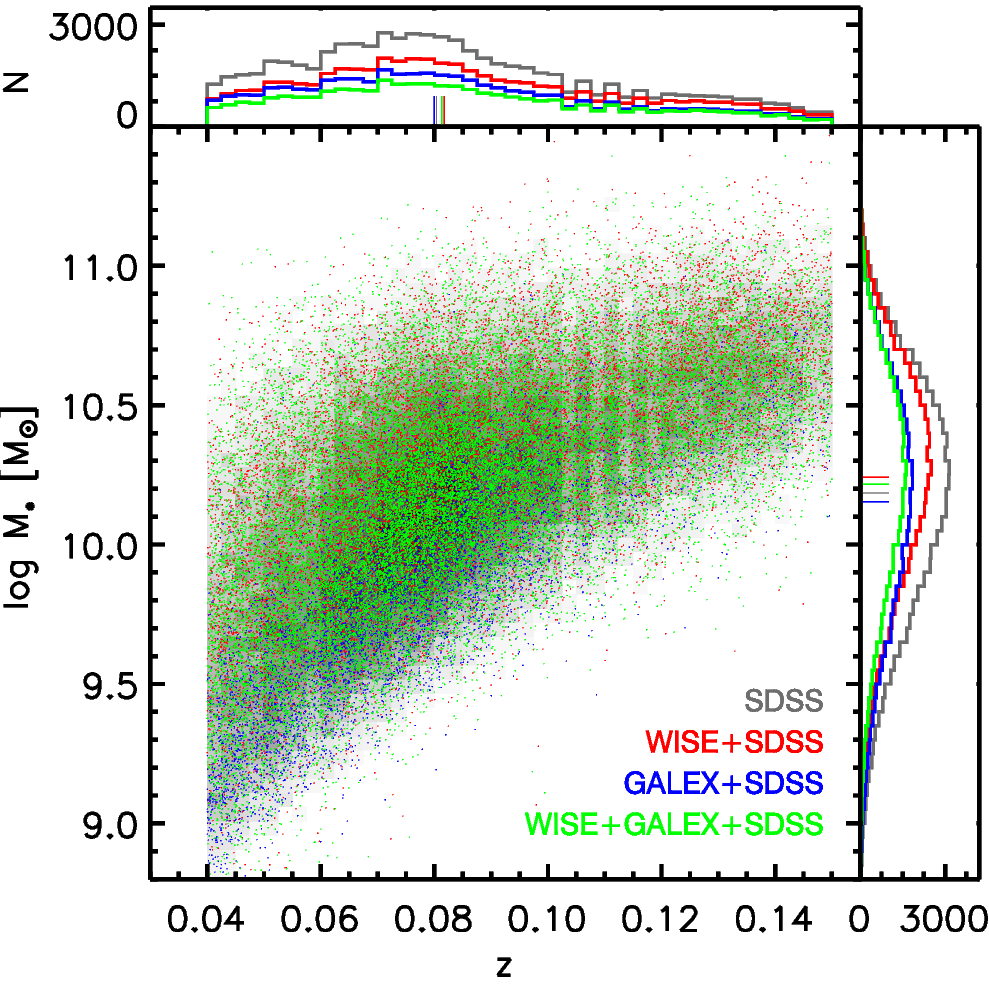}
\caption{Stellar mass as a function of redshift for local SFGs. The gray map shows the distribution of 67\,650 optically-selected SFGs from SDSS. The blue, red and green points mark the SDSS SFGs with detections in GALEX, WISE and GALEX+WISE observations, respectively. Their histograms are shown in the upper and right panel, marking the distribution of redshift and stellar mass respectively. The short solid lines represent the corresponding median value for different subsample.}
\label{fig:fig1}
\end{figure}

\section{SAMPLE SELECTION AND DATA}\label{sec:sec2}
\subsection{Sample Selection}\label{sec:sec2.1}

A sufficiently large sample of local SFGs is needed to systematically examine the relationships between IRX and other galaxy parameters. We select local galaxies with spectroscopic observations from the Sloan Digital Sky Survey Data Release 10 \citep[SDSS DR10,][]{Ahn2014}. 
We adopt the DR10 value-added catalog produced by the group from Max Planck Institute of Astrophysics and Johns Hopkins University (MPA/JHU)\footnote{\url{https://www.sdss3.org/dr10/spectro/galaxy_mpajhu.php}}, including spectroscopic line measurements as well as galaxy parameters derived  from the $u$, $g$, $r$, $i$ and $z$ imaging and spectroscopic data. The catalog contains 1\,477\,411 sources, and 841\,302 of them are from the SDSS Main Galaxy Sample (MGS) with  $r$$<$17.77 \citep{Strauss2002}.

We limit our sample to the redshift range 0.04$<$$z$$<$0.15 and remove sources with insecure redshift measurements ($Z\_WARNING >0$). The lower limit of redshift is chosen following \mbox{\citet{Kewley2005}} to ensure the fibre taking $>$20 per cent of total star light of a typical galaxy and minimize the potential differences between nuclear and global measurements in metallicity. We further get rid of galaxies with fibre coverage fraction $<0.2$, where the fibre fraction, parameterised by the fibre-to-total stellar mass ratio, is derived based on the MPA/JHU catalog.  The upper limit of $z$$<$0.15 is set to minimize evolutionary effects. We also get rid of the sources spectroscopically classified as a quasi-stellar object (QSO). The objects with multiple entries are excluded to keep our sample clean (we removed all of them). Here the multiple entries refer two or more sources in the catalog with angular separation $<5\arcsec$ and $\Delta z<0.001$. With an angular separation cut of $<5\arcsec$, some interacting galaxies would have been removed from our analysis. Since galaxy pairs with such close separation only compose of a very small fraction ($\sim 2$ per cent) of the whole galaxy population \citep[see table 2 of][]{Patton2008}, removing these galaxies will not affect our results. There are 409\,513 sources left after applying these selection cuts.

 We use the BPT diagnostic \citep{BPT1981} to separate SFGs and active galactic nuclei (AGN). This involves \oiii, H$\beta$, \nii\ and H$\alpha$.  Note that the measurement uncertainties on these line fluxes given in the MPA/JHU catalog are significantly underestimated. We adopt a factor of 2.473, 1.882, 1.566 and 2.039 for H$\alpha$, H$\beta$, \oiii\ and \nii, respectively, as recommended by the MPA/JHU team\footnote{\url{https://wwwmpa.mpa-garching.mpg.de/SDSS/DR7/raw_data.html}}. Following \citet{GB2010}, we adopt a threshold of signal-to-noise ratio (S/N) of 20, 3, 3 and 2 for the H$\alpha$, H$\beta$, \nii\ and \oiii\ emission lines. A cut of S/N$>$20 for H$\alpha$ ensures that the majority of sources have the four emission lines detected with sufficient S/N, minimizing the selection bias caused by the S/N cuts of individual emission lines. The S/N threshold for \oiii\ is decreased to 2 to avoid bias against weak \oiii\ emitters. This will not affect the BPT diagram diagnostic. In total 99\,054 objects remain after applying the emission line S/N cuts. Next we identify a clean sample of 84\,333 SFGs using the BPT selection criteria  adopted from \citet{Kauffmann2003}.

 We utilize the FUV and NUV data from the wide surveys by the Galaxy Evolution Explorer (GALEX) \citep{Martin2005} and the 22\,$\mu$m data from the Wide-field Infrared Survey Explorer (WISE) All Sky Survey \citep{Wright2010}. The cross-matching of SDSS targets with GALEX and WISE photometric catalogs has been done by \citet{Salim2016}. We cross-correlate our sample with the GALEX-SDSS-WISE Legacy Catalog from \citet{Salim2016} to obtain UV and IR photometry for our sample. Following \citet{Salim2016}, we use FUV fluxes measured at NUV positions ($FUV\_NCAT$) rather than from independent FUV detections, as they provide a more robust UV color. Then the WISE fluxes are taken from the unWISE catalog \citep{Lang2016}, which performed forced photometry on WISE images using SDSS centroids and profiles as priors.  It is worth noting that the prior-based photometry provides better flux measurements than the PSF photometry, particularly for sources that are blended in the WISE observations.
Limiting WISE 22$\mu m$ detections at the $>$2$\sigma$ level, we find that $\sim74$ per cent of our sample SFGs have 22$\mu m$ counterparts. The GALEX data are from the final data release (GR6/7). The NUV and FUV detections are limited above the 3$\sigma$ level to give a reliable estimate of UV luminosities. We will discuss the potential biases from these cuts in Section~\ref{sec:sec2.3}. Our final sample counts 32\,354 SFGs. Table~\ref{tab:tab1} lists the details of our sample selection.

\subsection{Measurements of Galaxy Parameters}\label{sec:sec2.2}

\subsubsection{Gas-Phase Metallicity}\label{sec:sec2.2.1}

The gas-phase metallicity in galaxies can be derived from strong line diagnostics, such as ([O\,{\small II}]$\lambda$3727 + [O\,{\small III}]$\lambda\lambda$4959,5007)/H$\beta$ (R23), \nii/H$\alpha$ (N2) or (\oiii/H$\beta$)/(\nii/H$\alpha$) (O3N2) \citep[e.g.][]{McGaugh1991, SB1994, KD2002, PP2004, Maiolino2008, Zahid2014a,Curti2017}. The systematic offsets between different metallicity indicators may reach up to a factor of 3 or even higher \citep[see][for a review]{KE2008}. The physical origin of this discrepancy is still not fully understood. The R23 calibration makes use of \oii\ which is prone to dust extinction and the correction for extinction relies on the estimate of dust obscuration.  Such correction might introduce a potential bias into the correlation between dust obscuration and gas-phase metallicity. Instead, N2 and O3N2 are less affected by dust extinction since only ratios of lines close to each other in wavelength are considered. Moreover, the emission lines in N2 and O3N2 are also used to identify SFGs from the BPT diagram. We can therefore measure a metallicity for our entire sample without introducing additional selection criteria. For a typical SFG in our sample, \nii\ and H$\alpha$ are usually much stronger than \oiii\ and H$\beta$. The typical measurement error of \nii/H$\alpha$ is smaller than that of \oiii/H$\beta$. We thus choose N2 to measure the metallicity for the entire sample. We tested that adoption of O3N2 metallicity diagnostic does not alter the conclusions presented in this paper.

We estimate the oxygen abundance (O/H) from N2 using the conversion formula given by \citet{PP2004} as
\begin{equation}\label{eq:eq1}
\zoh= 9.37 + 2.03\times {\rm N2} + 1.26\times {\rm N2}^2 + 0.32\times {\rm N2}^{3},
\end{equation}
where N2$=\log$(\nii/H$\alpha$). This relation is  valid over $-2.5<$N2$<-0.3$, corresponding to $7.17<\zoh<8.86$ \citep{PP2004}. In our analysis of metallicity, we exclude sample galaxies with N2 out of this range. For reference, Solar metallicity corresponds to an oxygen abundance of $\zoh=8.69$ \citep{Asplund2009}.  The metallicity errors account for measurement errors of line fluxes and calibration uncertainties. Benefiting from the high S/N limit on H$\alpha$, the majority of SFGs (> 99 per cent) in our sample have S/N>8 for [NII], much higher than the S/N=3 limit cut. The typical error is 6 per cent for N2, or 0.03\,dex for $\zoh$. The uncertainty of our metallicity estimate is dominated by the calibration uncertainty for N2, yielding 1$\sigma$=0.18\,dex in $\zoh$ \citep{PP2004}.

\subsubsection{Structural Parameters}\label{sec:sec2.2.2}

Galaxy structural parameters are taken from \citet{Simard2011}, who fit two-dimensional surface brightness profiles to the galaxy images for a sample of 1\,123\,718 galaxies from SDSS, providing the best-fitting single S\'{e}rsic model and the best-fitting two-component model (an exponential disc plus de Vaucouleurs bulge with S\'{e}rsic index $n_{\rm b}=4$) for each galaxy in $g$ and $r$, respectively. The semi-major axis half-light radius of the best-fitting $r$-band disc+bulge model is used as galaxy size ($R_{\rm e}$).  The axial ratio $b/a$ of the disc component from the best-fitting $r$-band disc+bulge model is adopted as the proxy of disc inclination ($b/a=cos(i)$ for a thin disc).  The typical error in $b/a$ is about 0.02. Galaxies lacking measurements of structural parameters are excluded from our analysis (see Table~\ref{tab:tab1}).

\subsubsection {Stellar Mass}\label{sec:sec2.2.3}

Stellar masses of our galaxies come from the MPA/JHU value-added catalog of SDSS DR10. The method to estimate the stellar masses is presented by \citet{Kauffmann2003} and \citet{Salim2007}, and the resulting quantities are consistent with those given by other works \citep{Taylor2011, Chang2015}. The median of the probability distribution is taken as the best estimate of stellar mass, and the uncertainty is derived from the 16th and 84th percentiles. We convert  the Kroupa IMF adopted in the MPA/JHA catalog to the Chabrier IMF through dividing stellar masses by a factor of 1.08 \citep{Madau2014}. 

\setcounter{table}{0}
\begin{table}
\centering
\caption{Summary of our sample selection.} \label{tab:tab1}
\begin{tabular}{@{}lrrr@{}}
\hline
 Selection Cut &  $N_{\rm gals}$ (remain) &  $F_{\rm R}$ (\%)$^{a}$  & $F_{\rm T}$ (\%)$^{b}$ \\
 \hline
 
initial sample                      & 1\,477\,411   &   -       &   -    \\ 
$r_{\rm petro}$$<$17.77             &    841\,302   &   43.06   &    43.06 \\
0.04$<$$z$$<$0.15                   &    461\,004   &   45.20   &    25.74\\
secure $z$                          &    460\,514   &    0.11   &    0.03\\
reject QSOs                         &    457\,504   &    0.65   &    0.20\\
single entry                        &    409\,513   &   10.49   &    3.25\\
S/N(H$\alpha$)$>$20                 &    101\,720   &   75.16   &    20.83\\
S/N(H$\beta$)$>$3                   &    101\,641   &    0.08   &    0.005\\
S/N(NII)$>$3                        &    101\,619   &    0.02   &    0.001\\
S/N(OIII)$>$2                       &     99\,054   &    2.52   &    0.17\\
BPT-selected SFGs                   &     84\,333   &   14.86   &    1.00\\
size/axial ratio                    &     83\,069   &    1.50   &    0.09\\
stellar mass                        &     83\,069   &    0.00   &    0.00\\
$f_{\rm fibre}$>0.2                 &     68\,464   &   17.58   &    0.99\\
$-2.5<N2<-0.4$                      &     67\,650   &    1.19   &    0.06\\
S/N(22$\mu$m)$>$2                   &     49\,944   &   26.17   &    1.20\\
S/N(FUV+NUV)$>$3                    &     32\,354   &   35.22   &    1.19\\
\hline
\multicolumn{4}{l}{$a$ -- percentage of remaining objects removed;} \\ 
\multicolumn{4}{l}{$b$ -- percentage of the initial parent sample removed.}\\ 

\end{tabular}

\end{table}

\subsubsection{IR \& UV Luminosity,  IR Excess and SFR}\label{sec:sec2.2.4}

We estimate the IR luminosity (8$-$1000\,$\mu$m) from WISE 22\,$\mu$m flux using a library of luminosity-dependent IR templates from \citet{CE2001}.  It has been shown that the IR luminosities derived from 22\,$\mu$m fluxes agree well with those estimated from Herschel/SPIRE 250, 350 and 500\,$\mu$m fluxes with a scatter of 0.07 dex \citep{Salim2016}.  The typical error of our IR luminosities is about 0.2\,dex, mainly attributed to the measurement errors of WISE 22\,$\mu$m fluxes as well as the calibration scatter.

The UV luminosity  (1216$-$3000\,\AA) is calculated from the integration of a galaxy SED template best fitting the observed FUV, NUV and $u$-band fluxes.  Here we firstly correct the FUV, NUV and SDSS $u$-band fluxes for  the Galactic extinction given in \citet{Schlegel1998} using the methods from \citet{Salim2016}.  
Following \cite{Salim2007}, we add modest calibration errors (from repeating observations) of 0.052, 0.026 and 0.02 mag to the FUV, NUV and $u$-band photometry, respectively. The FUV, NUV and $u$-band fluxes are fitted with a library of galaxy SED templates with an e-folding ($\tau$=1\,Gyr) history from \citet{BC2003} and the best-fitting template is used to derive the UV luminosity.  The typical error on $L_{\rm UV}$ is $\sim$15 per cent , mainly attributed to photometric errors.

As mentioned before, IRX refers to the ratio between the IR luminosity and UV luminosity ($\irx$), representing a measure of obscured over unobscured star formation and hence dust obscuration. The typical error on IRX is $\sim$0.2\,dex, contributed by the uncertainties of IR and UV luminosities. We caution that the measured UV luminosities may be subject to inclination effects whereas the IR luminosities are free of such a projection bias.  We account for this effect by introducing a $b/a$ dependence in our empirical IRX relation.

We calculate SFR following \citet{bell2005} and rescale to a \citet{Chabrier2003} IMF:
\begin{equation}\label{eq:eq2}
  {\rm SFR} = 1.09\times 10^{-10}\,( L_{\rm IR} + 2.2\,L_{\rm UV}),
\end{equation}
where $L_{\rm UV}$ and $L_{\rm IR}$ are given in units of $L_\odot$ with $L_\odot$=3.83$\times10^{33}$\,erg\,s$^{-1}$ and SFR is given in units of $M_\odot$ yr$^{-1}$ with $M_\odot$=1.99$\times10^{33}$\,g.

\subsection{Summary of the Sample}\label{sec:sec2.3}

Our goal is to examine the relationships of IRX with other galaxy parameters. This requires a sample spanning the full parameter space although not necessarily to high completeness. The criteria applied in our sample selection are summarized in Table~\ref{tab:tab1}. These selection steps are mostly related to the secure detections or reliable measurements for the considered galaxy observables. The last two columns of Table~\ref{tab:tab1} show the percentage of removed galaxies in that step, computed relative to the sample remaining from the previous step or relative to the initial parent sample, respectively. The latter provides a straightforward comparison of the importance of different cuts. We can see from Table~\ref{tab:tab1} that 67\,650 SFGs are selected from SDSS DR10 with 0.04$<$$z$$<$0.15 and stellar and structural parameters derived from optical observations. Figure~\ref{fig:fig1} shows stellar mass as a function of redshift for those optical-selected SFGs (gray scale). Of them, $\sim$74 per cent  of our sample SFGs have WISE 22$\mu m$ detections (red points), $\sim$59 per cent are detected in the FUV and NUV at the $>$3$\sigma$ level (blue points), and only $\sim$48 per cent have detections in both GALEX FUV+NUV and WISE 22\,$\mu$m observations (green points). Finally, there are 32\,354 SFGs having stellar mass ($M_\ast$), SFR, metallicity, $L_{\rm IR}$, $L_{\rm UV}$, IRX, $R_{\rm e}$ and axial ratio ($b/a$) available.  Our sample spans the full distribution of masses and redshifts of the underlying SDSS population without IR or UV criteria imposed, with completeness levels that are relatively flat across the diagram.  Using the Balmer decrement (H$\alpha$/H$\beta$) as an extinction indicator, we find that the galaxies with GALEX detection will bias against the more dusty galaxies. In contrast, requiring a WISE detection will bias against the less dusty galaxies instead. Since we focus on investigating the parameterised relations and still retain a dynamic range of nearly two orders of magnitude in IRX, such selection biases will not significantly affect our results. 

\begin{figure*}
\includegraphics[width=1\textwidth]{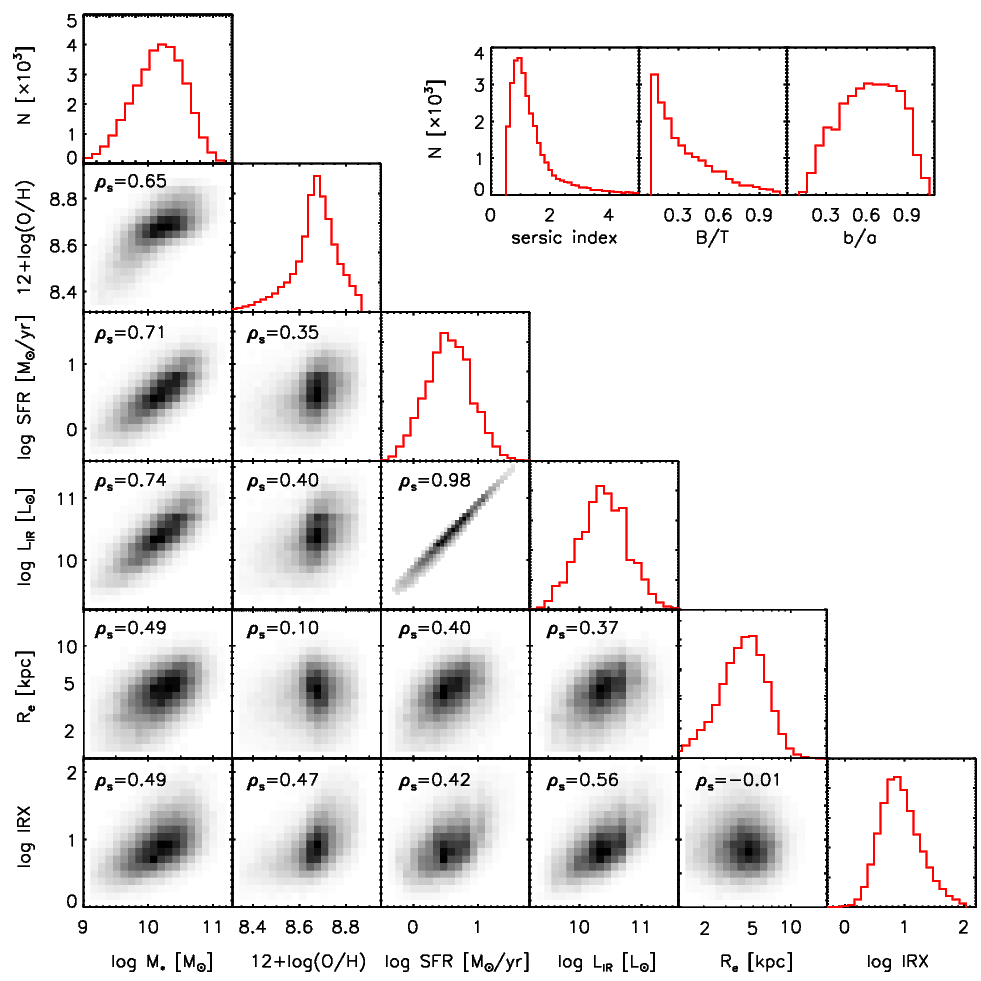}
\caption{Relationships between stellar mass, SFR, metallicity, the IR luminosity, galaxy size and IRX for our final sample of 32\,354 local SFGs. The upper panels show the histograms of these parameters.   The distribution of structural parameters is illustrated by the inset panels on the upper right.  Spearman's rank correlation coefficients $\rho_{\rm s}$ are given to show the strength of the correlation in each panel. It is clear that SFR, metallicity, IR luminosity, galaxy size are all strongly correlated with stellar mass. The correlation of IRX with the IR luminosity is actually stronger than that with stellar mass. The other panels show weak or no correlations when SFGs of different stellar masses are mixed together, suggesting that any correlations between these parameters reflect an indirect imprint of their joint dependence on stellar mass.}
\label{fig:fig2}
\end{figure*}

\section{THE DEPENDENCE OF IRX ON GALAXY PROPERTIES}\label{sec:sec3}

We examine the global relationships between galaxy parameters using our carefully-selected sample of local SFGs. Figure~\ref{fig:fig2} shows the distributions of $M_\ast$, SFR, $\zoh$, $L_{\rm IR}$, IRX, $R_{\rm e}$ and $b/a$.  We calculate Spearman's rank correlation coefficient ($\rho_{\rm s}$) for each diagram.  It can be seen that stellar mass is tightly correlated with SFR, metallicity, IR luminosity and galaxy size. These are well established scaling relations for SFGs. Other correlations in Figure~\ref{fig:fig2} are mostly driven by the primary scaling relations with mass in the first place. The tight link between SFR and the IR luminosity is because the IR luminosity dominates the bolometric luminosity (i.e., SFR) for our sample of SFGs.  Interestingly, the IR luminosity, as a measure of obscured SFR, shows a stronger correlation with stellar mass than SFR does.


It is not surprising that IRX is correlated with stellar mass, metallicity and SFR or IR luminosity in the sense that more massive SFGs tend to be higher in metallicity, SFR and dust obscuration. These correlations have been widely addressed in previous studies \citep{Cortese2006,Reddy2010,Whitaker2014,Whitaker2017}. The correlation of IRX with $L_{\rm IR}$ is the strongest  ($\rho_{\rm s}$=0.56), suggesting that IR luminosity is the most relevant to dust obscuration among these parameters. This may partially be due to the fact that the IRX directly contains IR luminosity and any uncertainties in $L_{\rm IR}$ may therefore introduce an artificial strengthening of the observed correlation between IRX and $L_{\rm IR}$.  We will come back to this issue in Section~\ref{sec:sec4}. Although IRX shows no correlation with galaxy size globally, IRX is correlated with IR luminosity surface density ($L_{\rm IR}/R_{\rm e}^2$), having $\rho_{\rm s}$=0.51 , respectively.  However, we find that IRX tightly correlates with $L_{\rm IR}/R_{\rm e}$ (i.e., IR line density), characterized by a $\rho_{\rm s}$ of 0.61. This is much stronger than the correlation with $L_{\rm IR}$ or $L_{\rm IR}/R_{\rm e}^2$, giving a clue that the IR line density may be mostly linked with the dust obscuration, and counterintuitively more so than the surface density.

\begin{figure*}
\includegraphics[width=1\textwidth]{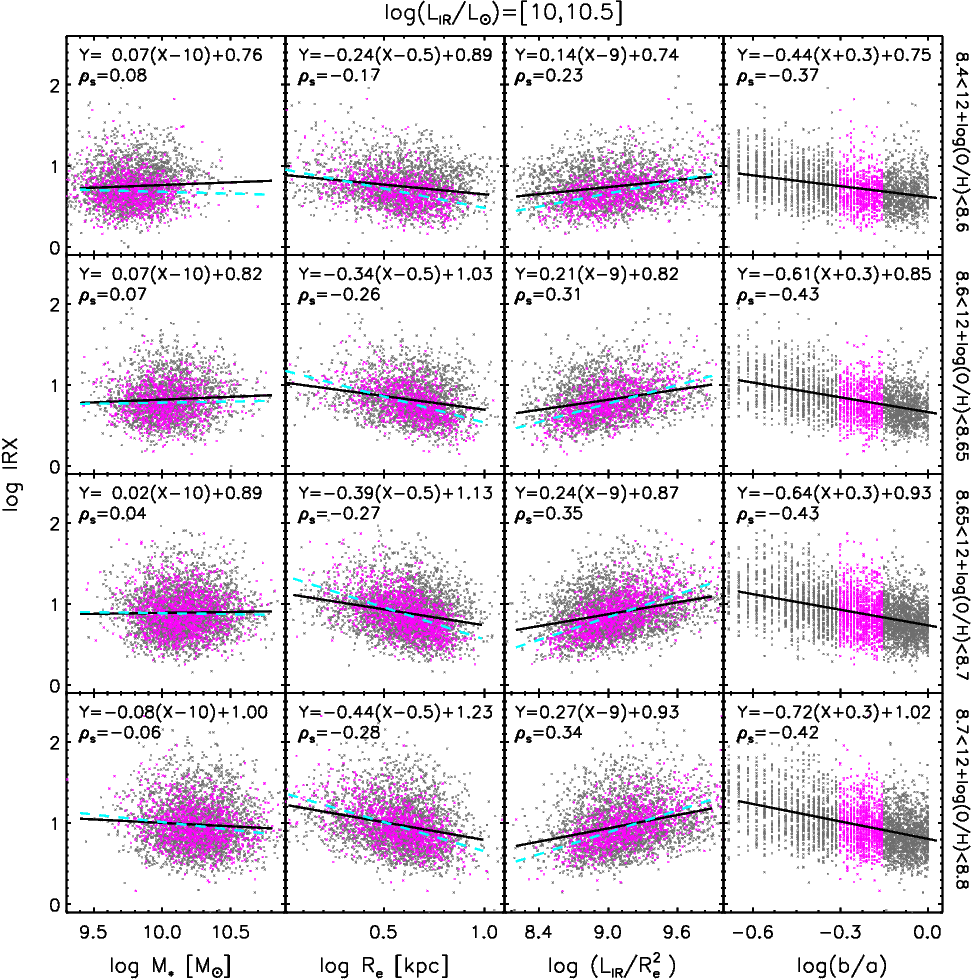}
\caption{Relationship between IRX and galaxy stellar mass, size,  $L_{\rm IR}/R_{\rm e}^2$, axial ratio (from left to right) for a subsample of local SFGs with 10$<$$\log L_{\rm IR}$$<$10.5 split into four metallicity bins over 8.4$<$$\zoh$$<$8.8 (from top to bottom). Spearman's rank correlation coefficients $\rho_{\rm s}$ are given to show the strength of the correlation in each panel. Black solid lines denote the best-fitting power-law relation and their best-fitting formulae are shown in each panel. The cyan dash lines refer to the best-fitting power-law relation to the data points with 0.5$<$$b/a$$<$0.7.  At fixed IR luminosities and metallicities, IRX shows no dependence on stellar mass. Instead,  IRX appears to correlate with both $R_{\rm e}$ and $L_{\rm IR}/R_{\rm e}^2$, and anti-correlate with $b/a$.  Note that with increasing metallicity the average IRX increases and the slopes become steeper.}
\label{fig:fig3}
\end{figure*}

\begin{table*}
 \centering
  \caption{The best-fitting power-law indices in the empirical IRX relation of Equation~\ref{eq:eq3} as a function of metallicity.}
  \label{tab:tab2}
  \begin{tabular}{@{}crcccc@{}}
  \hline
  Metallicity bin & $N_{\rm gals}$ &  $\alpha$ & $\beta$ &$\gamma$ &  $\delta$ \\
 \hline\hline
$8.30<\zoh<8.40$ &  656  & $ 0.62{\pm0.019}$ & $ 0.35{\pm0.020}$ & $0.42{\pm0.042}$ &    $0.42{\pm0.042}$ \\
$8.40<\zoh<8.50$ &  1738 & $ 0.69{\pm0.012}$ & $ 0.45{\pm0.013}$ & $0.53{\pm0.025}$ &    $0.61{\pm0.026}$ \\
$8.50<\zoh<8.60$ &  4315 & $ 0.77{\pm0.009}$ & $ 0.48{\pm0.009}$ & $0.61{\pm0.016}$ &    $0.78{\pm0.017}$ \\
$8.60<\zoh<8.65$ &  5434 & $ 0.88{\pm0.009}$ & $ 0.57{\pm0.009}$ & $0.69{\pm0.015}$ &    $0.84{\pm0.016}$ \\
$8.65<\zoh<8.70$ &  8454 & $ 1.00{\pm0.007}$ & $ 0.63{\pm0.007}$ & $0.87{\pm0.014}$ &    $0.96{\pm0.014}$ \\
$8.70<\zoh<8.75$ &  6107 & $ 1.03{\pm0.009}$ & $ 0.70{\pm0.008}$ & $0.95{\pm0.016}$ &    $1.08{\pm0.016}$ \\
$8.75<\zoh<8.80$ &  3307 & $ 1.02{\pm0.012}$ & $ 0.74{\pm0.011}$ & $0.89{\pm0.020}$ &    $1.03{\pm0.021}$ \\
$8.80<\zoh<8.86$ &  2120 & $ 1.07{\pm0.015}$ & $ 0.74{\pm0.013}$ & $0.85{\pm0.025}$ &    $1.02{\pm0.027}$ \\
\hline
\end{tabular}
\end{table*}

Next we examine the dependence of IRX on galaxy properties after removing the scaling relations related to galaxy stellar mass/SFR/luminosity/metallicity.  Since galaxy stellar mass, SFR and the IR luminosity are tightly correlated with each other and the IR luminosity is the most relevant to IRX, we contrast IRX with other parameters at a fixed IR luminosity and metallicity.  In practice, we select subsamples of our local SFGs in given IR luminosity bins, and then split each subsample into metallicity bins to see the correlations with different galaxy parameters in each bin.

We first focus on a subsample of 14\,305 SFGs with 10$\le$$\log( L_{\rm IR}/L_\odot)$$<$10.5 split into four metallicity bins over 8.4$<$$\zoh$$<$8.8.  Here the bin width of 0.5\,dex in $L_{\rm IR}$ is chosen as a compromise between a bin size large enough to allow robust statistics while minimizing the dynamic range of any $L_{\rm IR}$ dependencies within the bin. From the left panel of Figure~\ref{fig:figA1}, we can see that our full sample evenly spreads in the plot of $L_{\rm IR}$ versus $b/a$ for all four metallicity bins, suggesting that the split in metallicity does not induce a selection bias between $L_{\rm IR}$ and $b/a$.

Figure~\ref{fig:fig3} shows the relationships of IRX with $M_\ast$, $R_{\rm e}$, $L_{\rm IR}/R_{\rm e}^2$ and $b/a$ as a function of metallicity for the selected slice in IR luminosity. It is clear that IRX decreases with $R_{\rm e}$, increases with $L_{\rm IR}/R_{\rm e}^2$, and decreases with $b/a$, while IRX shows no dependence on stellar mass (with |$\rho_{\rm s}$|$<$0.08) for all four metallicity bins.   Moreover, the average IRX increases at increasing metallicity for a given set of ($M_\ast$, $L_{\rm IR}$, $R_{\rm e}$, $b/a$);  and the slopes of the best-fitting relations to the correlation of IRX with $R_{\rm e}$ and $L_{\rm IR}/R_{\rm e}^2$ and $b/a$, become steeper at increasing metallicity.

 The decrease of IRX with $b/a$ follows naturally from the range of viewing angles sampled, from edge-on for low $b/a$ to face-on for $b/a \sim 1$.  The observed $L_{\rm UV}$ from a galaxy seen edge-on will have suffered more extinction due to the larger dust column integrated along the line of sight. A negative correlation is also obtained when considering IRX as a function of $R_{\rm e}$, similar to that seen for $b/a$. On the other hand, we found that $R_{\rm e}$ is negatively correlated with $b/a$ (see the right panel of Figure~\ref{fig:figA1}). This is consistent with previous studies \citep{Huizinga1992,Mollenhoff2006,Yip2010}, which found that the smaller SFGs tend to be rounder. Therefore, the anti-correlations of IRX with $b/a$ and $R_{\rm e}$ shown in Figure~\ref{fig:fig3} might be coupled with each other to some extent. We investigate this issue by further limiting $b/a$ in the range of $0.5-0.7$ and revisiting the correlation between IRX and $R_{\rm e}$. As shown with magenta symbols in Figure~\ref{fig:fig3}, the dynamical range of $R_{\rm e}$ is not significantly affected, and the correlations of IRX with $R_{\rm e}$ or $L_{\rm IR}/R_{\rm e}^2$ become much tighter and steeper after removing the inclination dependence. The slopes of the best-fitting relations also become steeper with increasing metallicity. These results confirm that IRX does depend on the compactness of galaxies.

We emphasize that the correlation between IRX and stellar mass seen in Figure~\ref{fig:fig2} no longer holds after removing the dependence of IRX on IR luminosity and metallicity --- IRX is statistically almost unchanged while stellar mass spans a dynamical range of $\sim$1\,dex, as shown in  Figure~\ref{fig:fig3}.  The same conclusions can be obtained using the subsamples of SFGs with 9$\le$$\log( L_{\rm IR}/L_\odot)$$<$10 and 10.5$\le$$\log( L_{\rm IR}/L_\odot)$$<$11 (Figures~\ref{fig:figA2} and \ref{fig:figA3}). 
When instead controlling stellar mass and IR luminosity, as shown in Figure~\ref{fig:figA4}, correlations of IRX with $R_{\rm e}$, $L_{\rm IR}/R_{\rm e}^2$ and $b/a$ still exist, and an increasing trend with metallicity is obviously present. This is strong evidence that it is metallicity instead of stellar mass which acts as a driving quantity for IRX even though stellar mass and metallicity are globally correlated with each other. 
Similarly, it can be seen from Figure~\ref{fig:figA5} that IRX also correlates with $L_{\rm IR}$ over 1\,dex when controlling stellar mass and metallicity. We caution that the correlation between IRX and $L_{\rm IR}$ may partially be due to the fact that the IRX directly contains IR luminosity. Taken together, we conclude that IRX correlates more importantly with IR luminosity and metallicity than with stellar mass.

\begin{figure}
\includegraphics[width=0.48\textwidth]{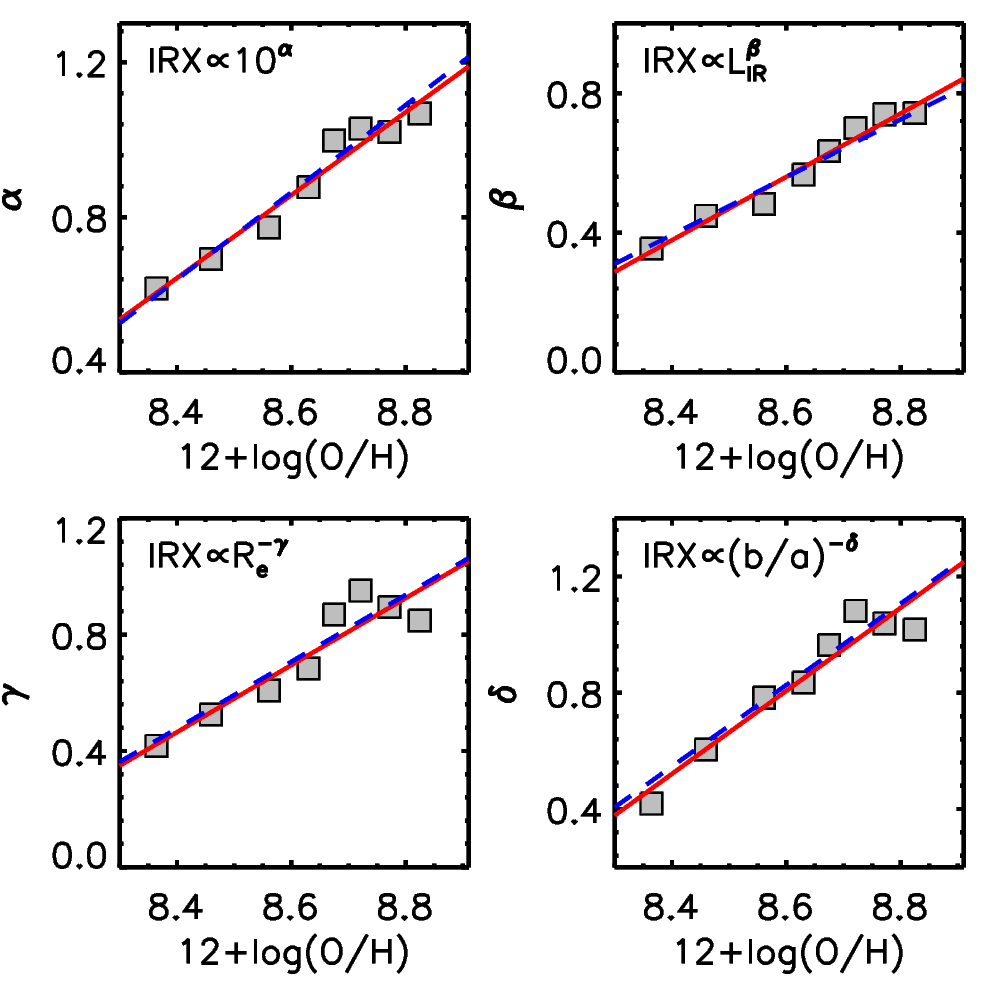}
\caption{Power-law indices $\alpha$,$\beta$, $\gamma$ and $\delta$ as a function of metallicity. Squares represent the best-fitting values given in Table~\ref{tab:tab2}. In each panel, the solid straight line gives the best fit to the data points using the least-squares method, and the dashed line represents the result from Equation~\ref{eq:eq5} fitting the full sample simultaneously. The two approaches yield consistent results. }
\label{fig:fig4}
\end{figure}

\section{THE EMPIRICAL IRX RELATION}\label{sec:sec4}

We have shown that IRX is dependent on IR luminosity (also SFR), size, metallicity and axial ratio among local SFGs. The anti-correlation with axial ratio is apparently attributed to the projection effect of disc SFGs impacting the observed UV luminosity.  These correlations imply that IRX is actually ruled by multiple variables.  We thus attempt to assess an empirical relation of IRX with  IR luminosity, size, metallicity and axial ratio by minimizing the scatter of the relation.

Considering that the majority of our sample lies within 0.2\,dex in metallicity, we firstly divide our sample into continuous metallicity bins over a range of 0.6\,dex and address IRX in relation to the other galaxy parameters in each metallicity bin. This will allow us to easily pinpoint the metallicity-sensitive changes. Motivated by the power-law relations shown in Figure~\ref{fig:fig3}, we assume that IRX obeys a formula
 \begin{equation}\label{eq:eq3}
IRX=10^\alpha\,(\frac{L_{\rm IR}}{10^{10}L_\odot})^{\beta}\,(\frac{R_{\rm e}}{\rm kpc})^{-\gamma}\,(b/a)^{-\delta},
\end{equation}
 where $\alpha$, $\beta$, $\gamma$ and $\delta$ are power-law indices to be determined via fitting to our sample. Here $b/a$ is included to separate the variation of IRX due to projection effects. The first term $10^\alpha$ simply records the normalization. Stellar mass is not included in the formula because IRX shows no dependence on stellar mass once controlling $L_{\rm IR}$ and metallicity, as shown in Figure~\ref{fig:fig3}. We verified this by including stellar mass in the fitting and present the results at the end of this section. The IDL package  MPFIT \citep{Markwardt2009} is used to fit the subsample of SFGs in a given metallicity bin with a $\chi^2$ minimization method. The fitting results are listed in Table~\ref{tab:tab2} for all eight metallicity bins. Note that the uncertainties for the best-fitting power-law indices are largely driven by the dispersion of data points, although only the uncertainties of IRX (typically $\sim$0.2\,dex) are counted in the fitting.

\begin{figure*}
\begin{center}
\includegraphics[width=1\textwidth]{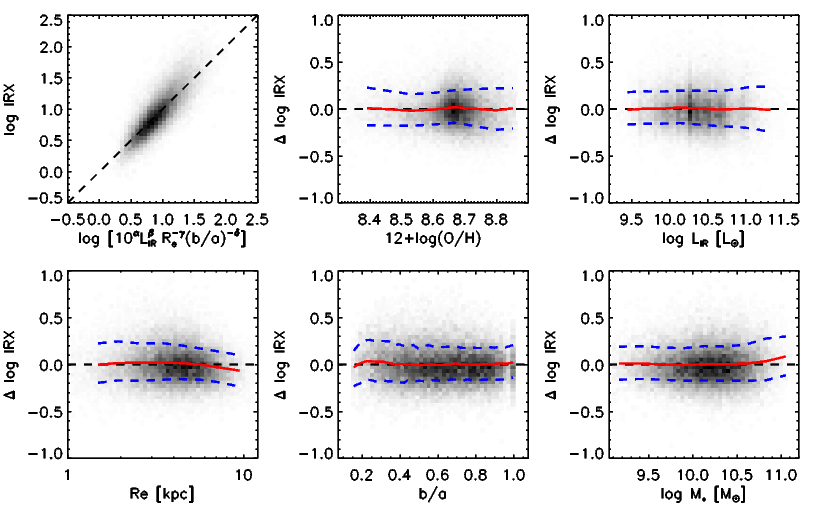}
\caption{Top-left: comparison of the predicted IRX from the best-fitting empirical relation with the observed IRX for our sample of local SFGs.  The remaining panels show the IRX residuals $\Delta \log IRX$ as a function of metallicity (top-middle), IR luminosity (top-right), galaxy size (bottom-left), axial ratio (bottom-middle) and stellar mass (bottom-right). The red and blue lines mark the median value and 1$\sigma$ dispersion of the data points. The black dashed-lines represent $\Delta \log IRX=0$.}
\label{fig:fig5}
\end{center}
\end{figure*}

It is clear from Table~\ref{tab:tab2} that all metallicity bins yield $\beta$, $\gamma$ and $\delta$ $>$ 0, meaning IRX increases with increasing IR luminosity, decreasing size and decreasing axial ratio (i.e., from face-on to edge-on). With high significance, we establish that these power-law indices increase with metallicity. At increasing metallicity over the range from 0.4 to 1.5 Solar metallicity, $\beta$ increases from 0.35 to 0.74, $\gamma$ from 0.42 to 0.85, and $\delta$ from 0.42 to 1.02. The normalization index $\alpha$ also increases with metallicity, indicating that more enriched SFGs feature higher dust obscuration.

Figure~\ref{fig:fig4} plots $\alpha$, $\beta$, $\gamma$ and $\delta$ as a function of metallicity. It can be seen that the power-law indices correlate with the logarithm of metallicity in a broadly linear manner. The best-fitting relations using the least-squares method are the following:
\begin{eqnarray}\label{eq:eq4}
& \alpha=1.07\log(Z/Z_\odot)+0.95, \nonumber \\
& \beta=0.91\log(Z/Z_\odot)+0.64, \nonumber \\
& \gamma=1.15\log(Z/Z_\odot)+0.80, \nonumber \\
& \delta=1.43\log(Z/Z_\odot)+0.94.
\end{eqnarray}
Here $\log(Z/Z_\odot)=\zoh-8.69$ in term of oxygen abundance and $Z_\odot$ refers to Solar metallicity. Those best-fitting relations are shown with red solid lines in Figure~\ref{fig:fig4}. From the best-fitting relations, we find $\gamma/\beta\sim1.25$. This indicates that it is $L_{\rm IR}/R_{\rm e}^{1.25}$ mostly regulating dust obscuration instead of the IR luminosity surface density ($\Sigma_{\rm IR}$=$L_{\rm IR}/R_{\rm e}^{2}$). This finding indeed challenges the interpretation of IR luminosity surface density as the main driver of dust obscuration \citep[e.g.][]{Desert1990, DH2002, Chanial2007}. We caution that the correlation between IRX and $L_{\rm IR}$ may be slightly biased by measurement uncertainties in $L_{\rm IR}$ because IRX itself contains IR luminosity.  We simply estimate this bias by arbitrarily adding scatter into $L_{\rm IR}$, finding that a reasonably large error of 0.2\,dex  will lead to an overestimate of the index of $IRX\propto L_{\rm IR}^{0.56}$ by 15 per cent.  Corrected for this bias,  $\gamma/\beta$ would become slightly larger as $\gamma/\beta\sim [1.3 - 1.5]$, although the actual correction is hard to quantify due to the dependence of both indices on metallicity.   We will discuss it in Section~\ref{sec:sec6.3}.

 It is worth noting that these relations remain almost unchanged when we carry out a single fit to all SFGs over the full metallicity range with the power-law indices in Equation~\ref{eq:eq3} set as a function of metallicity. We parameterise the dependence on metallicity as
\begin{equation}\label{eq:eq5}
 X=c_X \log(Z/Z_\odot)+d_X, 
\end{equation}
where $X$ represents $\alpha$, $\beta$, $\gamma$ or $\delta$, and $c_X$ and $d_X$ are the corresponding coefficients. The best-fitting relations are shown with the blue dashed lines for comparison in Figure~\ref{fig:fig4}. We conclude that fitting all SFGs in our sample simultaneously (dashed lines) yields consistent results with those obtained by fitting to individual metallicity bins (solid lines). We choose the latter as our referenced IRX relation for the sake of not biasing against the metal-poor galaxies. Taken together, Equation~\ref{eq:eq3} in conjunction with Equation~\ref{eq:eq4} presents IRX as a function of metallicity, IR luminosity, galaxy size and axial ratio. We name it the empirical IRX relation.

Figure~\ref{fig:fig4} reveals that the values of $\beta$, $\gamma$ and $\delta$ become smaller for lower metallicity, meaning that the dependence of IRX on $L_{\rm IR}$, $R_{\rm e}$ and $b/a$ becomes weaker. These results denote that gas-phase metallicity is a key quantity for regulating dust obscuration.

We now examine the scatter of our galaxies around the empirical IRX relation. First we compare the observed IRX against the predicted IRX from the best-fitting empirical relation for our sample of local SFGs in the first panel of Figure~\ref{fig:fig5}.  We point out the prediction is made given a set of inputs ($\zoh$, $L_{\rm IR}$, $R_{\rm e}$ and $b/a$) using Equation~\ref{eq:eq3} + Equation~\ref{eq:eq4} for every SFG in our sample.  The predicted IRX agrees well with the observed value. For describing the scatter, we define the IRX residual as as $\Delta\,\log IRX=\log IRX_{\rm observed}-\log IRX_{\rm model}$, The $\Delta\,\log IRX$ distribution can be well fitted with a Gaussian profile of $\sigma=$0.17\,dex.  In Figure~\ref{fig:figA6}, we show the distribution of IRX residuals in five $IRX_{\rm model}$ bins. It is clear that  the dispersion increases with $IRX_{\rm model}$ over the range of $\sigma$=[0.15,0.24]\,dex.

The IRX residual is also shown as a function of metallicity, IR luminosity, galaxy size, axial ratio and stellar mass, respectively, in the remaining panels of Figure~\ref{fig:fig5}. One can see that $\Delta\,\log IRX$ does not correlate with these quantities used in the empirical IRX relation, confirming that the power-law indices of $\alpha$, $\beta$, $\gamma$ and $\delta$ are properly determined for the empirical IRX relation.

It is worthwhile coming back to the point whether stellar mass should be included as a component of the empirical IRX relation.  We ignored it in our fitting because IRX exhibits no dependence on stellar mass when IR luminosity and metallicity are fixed (see Figure~\ref{fig:fig3}). As expected, the IRX residuals presented in Figure~\ref{fig:fig5} are flat with respect to galaxy stellar mass. This confirms that the input galaxy parameters adopted in Equation~\ref{eq:eq3} fully account for the variability of dust obscuration. For a fair check, we additionally add a free term relying on stellar mass in Equation~\ref{eq:eq3} as
 \begin{equation}\label{eq:eq6}
IRX=10^\alpha (\frac{L_{\rm IR}}{10^{10}L_\odot})^{\beta}(\frac{R_{\rm e}}{\rm kpc})^{-\gamma}(b/a)^{-\delta} (\frac{M_\ast}{10^{10}M_\odot})^{\epsilon},
\end{equation}
 and repeat the fitting procedure. The power-law indices are adjusted to best reproduce the observed IRX for the subsample of SFGs of each metallicity bin. The fitting results are shown in Figure~\ref{fig:figA7}. Our fitting yields values for $\epsilon$ spanning [$-$0.08, 0.17], consistent with zero within the uncertainties. The other power-law indices $\alpha$, $\beta$, $\gamma$ and $\delta$, are almost unchanged with stellar mass included in the fitting. We see this result as strong evidence that the variability of IRX is mainly associated with IR luminosity, metallicity, size and axial ratio, and no explicit dependence on stellar mass is required.

 \begin{figure*}
\includegraphics[width=1\textwidth]{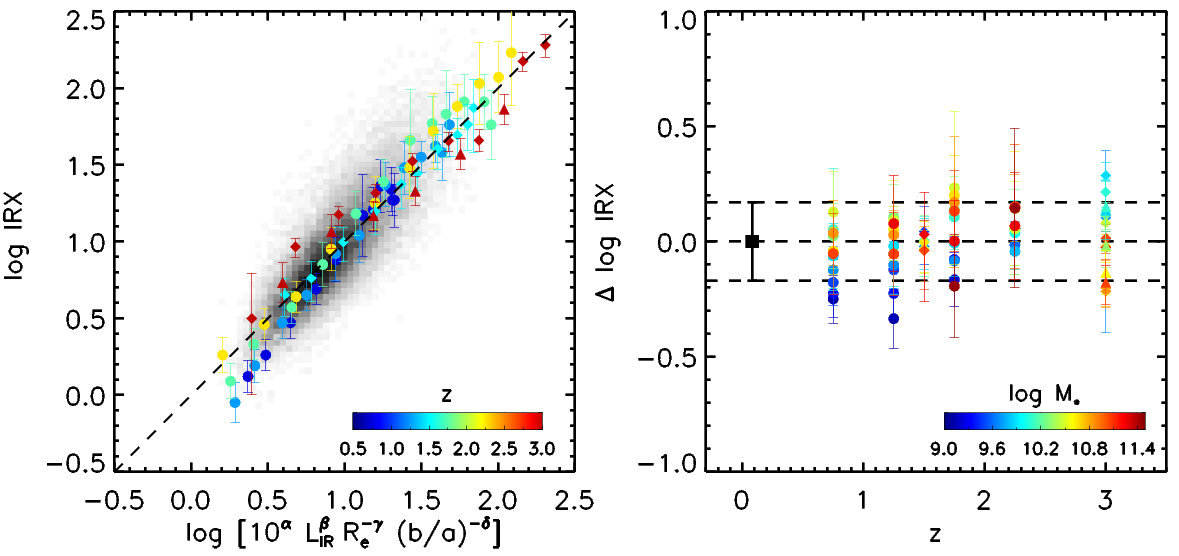}
\caption{Left panel: comparison of the observed IRX with the inferred IRX from our empirical IRX relation for our sample of local SFGs (grey points) and for a population-averaged census of high-$z$ SFGs (solid symbols color-coded with redshift). Right panel: IRX residuals as a function of redshift. The black square and its error bars (and the corresponding dashed lines) are the zero value and 1$\sigma$ dispersion ($\sim$ 0.17\,dex) for our sample of local SFGs. The high-$z$ symbols are similar to the left panel but color-coded with stellar mass instead. It is clear that distant SFGs follow the local SFGs within 1$\sigma$ dispersion along the empirical IRX relation.}
\label{fig:fig6}
\end{figure*}

\section{THE IRX RELATION FOR HIGH REDSHIFT GALAXIES} \label{sec:sec5}

It is interesting to see if distant SFGs obey the same empirical IRX relation as local SFGs. We collect the measurements of average IRX and IR luminosity for SFGs in different stellar mass and redshift bins in the literature, including \citet{Whitaker2014} for 0.5$<$$z$$<$2.5, \citet{Heinis2014} for $z$=1.5 and $z$=3, \citet{AM2016} for $z$=3. We derive the average gas-phase metallicity at a given stellar mass and redshift from the empirical mass-metallicity relations given by \citet{Genzel2015}, which combine the mass-metallicity relations at different redshifts presented by \citet{Erb2006}, \citet{Maiolino2008}, \citet{Zahid2014a} and \citet{Wuyts2014}. All the metallicity values have been calibrated to match the \citet{PP2004} N2-based metallicity using the formulae given in \citet{KE2008}. The average size of distant SFGs is estimated from the mass-size relations presented by \citet{VDW2014}. We also calculate the mean axial ratio of the 3D-HST sample \citep{Skelton2014} at different redshifts and apply this result to all subpopulations of distant SFGs. To derive the average axial ratio, we only consider the galaxy has good S\'{e}rsic profile fitting and a UVJ type of star-forming. In the left panel of Figure~\ref{fig:fig6}, we show the average IRX versus the inferred IRX based on the average properties using color-coded solid symbols (color-coded with redshifts) for high-$z$ SFGs divided into different stellar mass and redshift bins. The gray scale are the local SFGs from our sample.

We can see that the agreement between local and distant SFGs is surprisingly good. The averaged data points of subpopulations of high-$z$ SFGs lie on the empirical IRX relation of local SFGs.  The right panel of Figure~\ref{fig:fig6} presents the IRX residuals as a function of redshift. The color coding now represents stellar mass. No correlation between IRX residuals and stellar mass or redshift is observed. We conclude that the empirical IRX relation derived from local SFGs still holds for distant SFGs up to $z$=3.  Therefore, we claim that it is a universal relation of dust obscuration.

 \begin{figure*}
\includegraphics[width=1\textwidth]{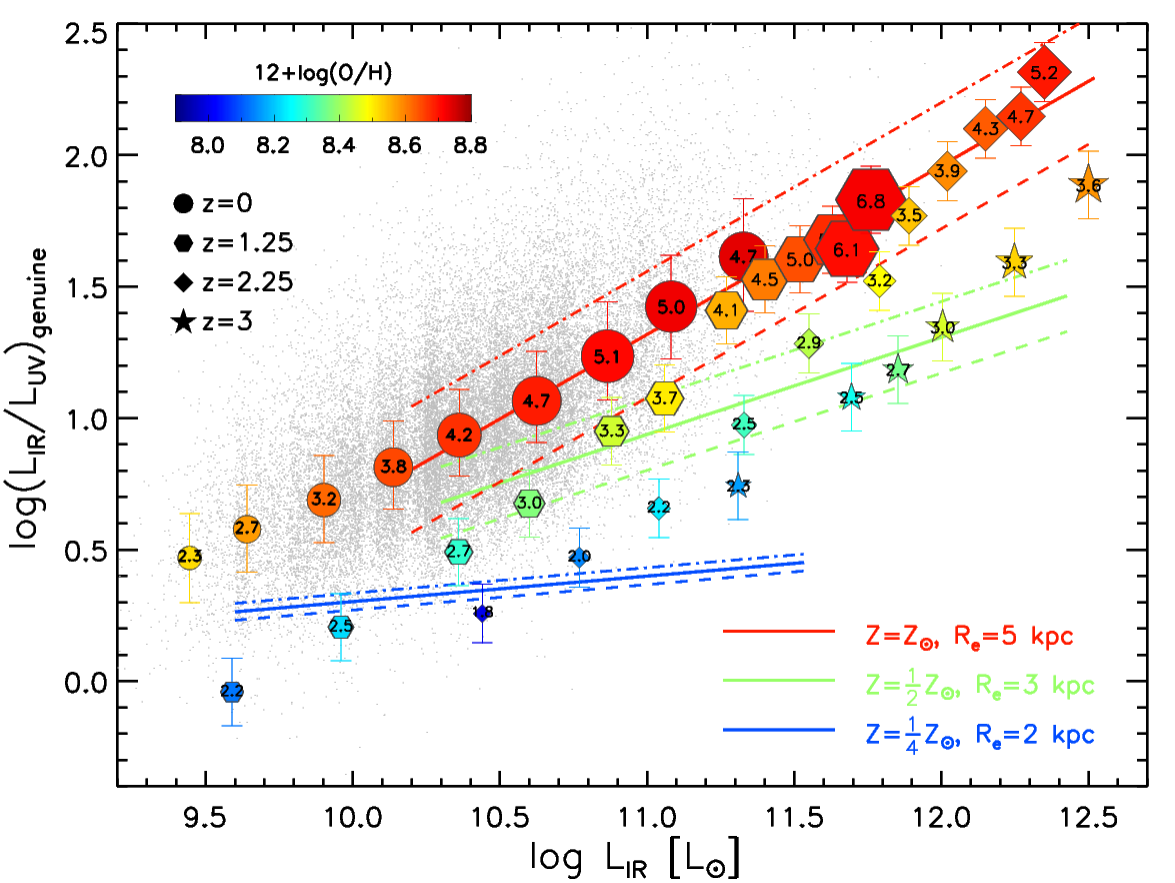}
\caption{The genuine IRX as a function of IR luminosity, metallicity and galaxy size. The gray points are our local 32\,354 SFGs and the solid circles with error-bars show the median value and 25-75 per cent dispersion in different $L_{\rm IR}$ bins. The hexagon, diamond and star symbols mark the typical SFGs of different IR luminosities at $z$=1.25, 2.25 and 3 respectively. The symbol size is proportional to the typical $R_{\rm e}$ (shown on the symbols in units of kpc) in the given subpopulation. The three solid lines represent the universal IRX relation at given metallicity and $R_{\rm e}$. The dashed and dash-dotted lines parallel to the solid line represent the relation with the sizes offset by $\pm$0.3\,dex (dashed lines have larger sizes). Both the symbols and lines are color-coded with the gas-phase metallicity shown by the color bar. }
\label{fig:fig7}
\end{figure*}

\section{Discussion}\label{sec:sec6}

Dust obscuration has been found to broadly correlate with some major physical, chemical and structural parameters in star-forming galaxies. The causal connection between these correlations remains to be understood. Our finding of a universal relation of dust obscuration reveals that IR luminosity, galaxy size, metallicity and axial ratio jointly delineate dust obscuration in both local and distant SFGs. Equations~\ref{eq:eq3} and \ref{eq:eq4} define how IRX quantitatively depend on these four parameters. From this relation, one can draw the following conclusions: 1) the inclination of SFGs affects IRX inducing an anti-correlation with $b/a$; 2) galaxy stellar mass is not a direct driver for dust obscuration; 3) IRX correlates more strongly with $L_{\rm IR}/R_{\rm e}^{[1.3 - 1.5]}$ rather than with $L_{\rm IR}$ surface density; 4) metallicity directly increases IRX and controls the power-law indices for IR luminosity, galaxy size and axial ratio. We discuss the implications of the universal relation of dust obscuration below.

\subsection{Enhanced IRX for more inclined systems}

IRX refers to the ratio of total luminosities at IR and UV wavelengths. While the IR luminosity can reasonably be thought of as isotropic, this is not the case for the observed UV light which is very sensitive to extinction by dust along the line of sight. For a disc, the optical depth dramatically changes with inclination and the observed UV luminosity is thus significantly biased compared to what would be measured integrating the emission over $4\pi$ steradians. The inclination-dependent deviation disperses the correlations of the genuine dust obscuration with other physical parameters. In our fitting, $(b/a)^{-\delta}$ is included to separate this projection effect from the other dependencies on intrinsic galaxy properties which encode actual physical variations in dust content or spatial distribution (where involving size).

The impact of projection effects on the observed IRX becomes larger with increasing metallicity. From the universal relation, IRX increases with decreasing $b/a$ satisfying a power-law with the power-law index $\delta$=[0.96,0.42] at [1, 1/2] Solar metallicity. This trend is consistent with results presented by \citet{Xiao2012} and \citet{Li2019}, who use the Balmer decrement as a measure of dust obscuration instead. Projection effects account for a large amount of variability of the observed IRX among SFGs of similar IR luminosities and metallicities, especially in the high-metallicity regime.

The universal IRX relation tells us that the observed UV luminosity is attenuated by the inclination as $L_{\rm UV}\sim(b/a)^{\delta}$. The optical depth for a galaxy with a homogeneous dust-star mixture geometry is given by $\tau$=$\tau_{\rm f}/\cos(i)=\tau_{\rm f}(b/a)^{-1}$, where $i$ is the inclination angle and $\tau_{\rm f}$ represents the optical depth when viewed face-on. At high metallicity, we do find that IRX is nearly inversely proportional to $b/a$ with $\delta\sim1$. However, at low metallicity, we obtain a $\delta=0.42$ and the anti-correlation becomes much flatter. This is consistent with the results given by \citet{Xiao2012}, in which a flattening correlation between dust obscuration and axial ratio is also favored at low metallicity. Considering the fact that the low-metallicity SFGs are usually less massive and tend to be more spheroidal in morphology \citep{PS2008}, the axis ratio is no longer driven mostly by inclination and therefore the IRX becomes less sensitive to the changing of axis ratio at low metallicity. 

We calculate that the genuine UV luminosity, defined by integrating the UV radiation emerging from a galaxy over $4\pi$ steradians, approximately equals the observed one at $b/a\sim0.5$. Removing the inclination-induced bias in IRX, we plot in Figure~\ref{fig:fig7} the genuine IRX as a function of IR luminosity, metallicity and galaxy size as derived from the universal IRX relation. Both local and distant SFGs are shown, split into $L_{\rm IR}$ bins. Note that the inclination correction is done for local individual SFGs but not for high-$z$ data points, since the ensemble average approach already averages over all (random) viewing angles. It becomes clear that the genuine IRX is jointly shaped by the three parameters following the universal relation. Solid lines represent the genuine IRX - $L_{\rm IR}$ relation for pairs of (typical) metallicity and size.  The dashed and dash-dotted lines illustrate how the relation changes when adjusting the galaxy size by a factor of 2 (dashed lines have larger size). At low metallicity, IRX significantly decreases and the correlation between IRX and $L_{\rm IR}$ becomes flatter. At the same time, IRX becomes less sensitive to galaxy size at low metallicity. Particularly, IRX becomes almost independent on galaxy size at 1/4 Solar metallicity. Noted that the relationship of 1/4 Solar metallicity shown in Figure~\ref{fig:fig7} is an extrapolation, out of the dynamical range of our local sample.

\subsection{IRX does not depend on stellar mass}

The mass-IRX relation has been widely examined over a wide redshift range \citep{Reddy2010,Heinis2014,Whitaker2014,AM2016}, finding that it evolves little out to $z$=3 at the low-mass end of $M_\ast$$<$10$^{10.5}\,M_\odot$ in contrast to a rapid evolution at the high-mass end \citep{Whitaker2014,Leslie2018}. Some authors concluded from this that galaxy stellar mass is the most fundamental parameter determining the dust obscuration in SFGs \citep{GB2010,Zahid2014b}.

Generally speaking, more massive SFGs have a higher SFR and a larger amount of dust in the ISM formed in the past, and tend to have a higher IRX.  However, we find that four free parameters $L_{\rm IR}$, $R_{\rm e}$, $Z$ and $b/a$ are sufficient to fit IRX and residuals from the universal relation show no dependence on stellar mass. The inclusion of stellar mass as a free parameter in our fitting gives $IRX\propto M_\ast^{[-0.08, 0.17]}$, suggestive of no correlation within the uncertainties. Studying the Balmer decrement in a sample of $\sim 1000$ nearby SFGs with integral-field spectroscopy, \citet{Li2019} come to the same conclusion, that trends of (spatially resolved) extinction with various local and global galaxy quantities can be reproduced by a model that lacks an intrinsic dependence on galaxy stellar mass.  Moreover, as shown in Figure~\ref{fig:fig6}, the high-$z$ SFGs of different stellar masses all follow the universal IRX relation derived from the nearby SFG population, further confirming that exclusion of stellar mass in our fitting works for distant SFGs as well.

The bolometric luminosity (UV+IR) of a star-forming galaxy is vastly dominated by young stars and can thus be seen as a good SFR indicator \citep{Kennicutt2012}. The dust obscuration tracer IRX is intimately related to the star formation process as well.  Dust in the birth clouds around young stars and in the surrounding ISM absorbs the UV radiation and re-emits into the IR. The overall dust obscuration of the galaxy in general depends on the dust column density averaged over the galactic scale. The latter relies on the gas density and gas-phase metallicity. The gas density on its turn is what sets the star formation surface density (i.e., the K-S Law).  While none of these quantities are directly linked with old stellar populations, SFR, metallicity, and galaxy size are all globally correlated with stellar mass. The observed mass-IRX relation can hence be considered the natural product of the universal relation of dust obscuration together with the scaling relations of stellar mass.  We thus conclude that stellar mass is not a causal parameter that directly controls the dust obscuration in an SFG.

\subsection{The term $L_{\rm IR}/R_{\rm e}^{[1.3 - 1.5]}$ and implications on the dust-star geometry} \label{sec:sec6.3}

One can wonder why IRX depends more strongly on $L_{\rm IR}/R_{\rm e}^{[1.3 - 1.5]}$ rather than on $\Sigma_{\rm IR}$ (=$L_{\rm IR}/R_{\rm e}^{2}$). The latter has been shown to tightly correlate with dust temperature among SFGs that are mostly discs \citep{Desert1990,Chanial2007} and with the Balmer decrement at a spatially resolved level modulo secondary dependencies on, e.g., metallicity and H$\alpha$ equivalent width \citep{Li2019}.  Note that the actual $\Sigma_{\rm IR}$ should be measured with the half-IR radius $R_{\rm IR}$ rather than the optical $R_{\rm e}$ which probes the spatial extent of the bulk of the stars, not of star formation.  We will come back to this point later.

Generally speaking, dust obscuration relies not only on the dust content but also on the geometry of the dust-star mixture. We consider two representative model geometries. One is a uniform foreground screen of dust. The intrinsic UV emission is attenuated by a factor of $e^{-\tau}$, where $\tau$ is the total optical depth of the obscuration material. The total optical depth in a galaxy is expected to be proportional to the dust column density $N_{\rm dust}$ along the line of sight (or the dust surface density $\Sigma_{\rm dust}$ when the galaxy is seen face-on), as $\tau \propto N_{\rm dust}\propto \Sigma_{\rm dust}$. We quickly dismissed such geometry, since it predicts that IRX increases with dust surface density exponentially. This is unrealistic, e.g., if the dust surface density increases by a factor of 10, the IRX will dramatically increase by a factor of $\sim$8000.

We then consider a more realistic geometry where dust and stars are homogeneously mixed. In such a geometry, the intrinsic UV emission is attenuated by a factor of  $(1-e^{-\tau})/\tau$ instead \citep{Schreiber2001,Wuyts2011}. This will suggest that
 \begin{equation}\label{eq:eq7}
IRX=2.2\,\left(\frac{\tau}{1-e^{-\tau}}-1\right)
   =2.2\,\left[\frac{\tau}{2},\tau\right]
   \propto \tau,
\end{equation}
 where 2.2 is the correction for the light emitted long-ward of 3000\,\AA\  and short-ward of 1216\,\AA\ by young stars \citep{bell2005} and Equation~\ref{eq:eq7} approaches $IRX \propto \tau$ within a scatter of only 0.3\,dex. Since the gas-phase metallicity is proportional to the dust-to-gas ratio, $\Sigma _{\rm dust}$ should follow the metallicity times the gas surface density. Using the K-S Law $\Sigma_{\rm SFR}\propto \Sigma_{\rm gas}^n$, we obtain
\begin{equation}\label{eq:eq8}
IRX\propto \Sigma_{\rm dust}\propto \left(\frac{Z}{Z_{\odot}}\right)\times({\rm SFR}/R_{\rm e}^2)^\frac{1}{n}
\end{equation}
with $n=1.42$ (or $1/n=0.7$) \citep{Kennicutt98}.  For a comparison with Equation~\ref{eq:eq8}, we simplify our universal IRX relation given in Equation~\ref{eq:eq3} and \ref{eq:eq4} into
\begin{equation}\label{eq:eq9}
IRX\propto \left(\frac{Z}{Z_{\odot}}\right)^{1.07}\times({\rm SFR}/R_{\rm e}^{1.25})^\beta,
\end{equation}
where $\beta=0.91\log(Z/Z_\odot)+0.64$. Here, we have replaced $L_{\rm IR}$ with SFR because $L_{\rm IR}$ approximately mirrors SFR at $IRX\gtrsim 3$ for massive SFGs of high metallicity. We can see that the values of $\beta$ and $1/n$ are nearly identical.

We realize that the size of $R_{\rm e}$ may differ from the radius $R_{\rm IR}$ enclosing half of all star formation \citep{Bianchi2007,Smith2016,Casasola2017}. The size of molecular gas that hosts star formation is usually smaller than the size of stars in massive galaxies. A relationship of $R_{\rm IR}\propto R_{\rm e}^{0.6}$ is required to interpret our universal IRX relation. Considering the bias induced by uncertainties of $L_{\rm IR}$ to the IRX-$L_{\rm IR}$ relation (i.e., replacing 1.25  with $[1.3 - 1.5]$), the difference between $R_{\rm IR}$ and $R_{\rm e}$ would become smaller, following $R_{\rm IR}\propto R_{\rm e}^{[0.6 - 0.75]}$. Correcting for such a difference between the IR size and the stellar size, our universal IRX relation would be in an excellent agreement with the inference from the K-S Law together with the geometry of a homogeneous dust-star mixture. We argue that this is the case at least for the local SFGs of high metallicity.

The inferred IRX from the K-S Law increasingly overestimates the actual observed IRX at decreasing metallicity. \citet{Wuyts2011} noticed this discrepancy that the inferred IRX through the K-S law plus a homogeneous mixture of dust+stars globally agrees with the observations at $z=0$ but becomes systematically higher at increasing redshift. The high-$z$ galaxies have a lower metallicity than their local counterparts, but different star-dust geometries may contribute to the observed trends as well. In fact, modeling the spatially resolved Balmer decrement distribution and its relation to other local and global properties, \citet{Li2019} conclude that generally a more complex dust geometry consisting of a homogeneous mixture and (clumpy) foreground screen component is needed, even in nearby galaxies, to adequately describe the observed extinction patterns.

From our universal IRX relation the power-law index $\beta$ drops to 0.35 at a half of Solar metallicity. \citet{Krumholz2009} pointed out that the star formation efficiency rapidly declines in the low metallicity SFGs, suggesting that $n$ in the K-S Law would increase in the low-metallicity regime. Such a tendency is consistent with our findings. This is associated with the fact that the threshold for conversion from atomic to molecular gas is observed (and theoretically understood) to be a function of metallicity \citep{Krumholz2009,Sternberg2014,Schruba2018}.

On the other hand, low-metallicity SFGs tend to be more spheroidal in morphology and systematically lower in stellar mass \citep{PS2008}. Our sample of local SFGs show a lack of low-metallicity systems with $b/a<0.3$, supporting that they are unlikely dominated by the rotating discs. The low-mass and low-metallicity SFG population is dominated by young stellar populations and unlikely to present a size discrepancy between old and young populations. We thus argue that the dust-star geometry in low-metallicity SFGs must differ from the disc geometry of a homogeneous mixture for the high-metallicity SFGs and yield an optical depth of $\tau$ insensitive to the galaxy size. Still, more efforts are needed to understand the reasons for our finding that the genuine IRX is actually a power-law function of $L_{\rm IR}/R_{\rm e}^{[1.3 - 1.5]}$.

\subsection{Systematics Coupled with Metallicity}

 As shown in Equation~\ref{eq:eq4}, the four power-law indices all change with metallicity, revealing that the gas-phase metallicity is the key to setting dust content and hence obscuration as well as its correlations with other parameters.  As shown in Figure~\ref{fig:fig7}, at decreasing metallicity IRX becomes less sensitive to either $L_{\rm IR}$ or $R_{\rm e}$. The inclination-induced bias to the genuine IRX rapidly declines. For the metal-poor SFGs of $Z<0.3\,Z_\odot$, the bias is negligible.

The normalization of the universal IRX relation increases with the gas-phase metallicity. It is worth noting that the normalization index $\alpha$  is measured at the references $L_{\rm IR}=10^{10}\,L_\odot$ and $R_{\rm e}=1$\,kpc in the empirical relation given by Equation~\ref{eq:eq3}. Since the power-law indices $\beta$, $\gamma$ and $\delta$ are all dependent on metallicity, the normalization index $\alpha$ would change if the references in $L_{\rm IR}$ and $R_{\rm e}$ change. This behavior is also illustrated in Figure~\ref{fig:fig7}. At increasing the metallicity over the range from 1/4 to 1 Solar metallicity, the IRX increases more at higher IR luminosity end.

We argue that the physics underpinning the best-fitting relations given in Equation~\ref{eq:eq4} likely comprises a coupling between the metallicity, gas fraction, compactness of star-forming regions, dust-star geometry, metal-to-dust ratio and size distribution of dust grains.  Our results suggest that these systematics coupled with metallicity should also be present among distant SFGs. It would be interesting to see whether models tracing the formation, destruction and spatial distribution of dust more from first principles, and rooted in a galaxy formation framework \citep[e.g.,][]{Feldmann2015,Popping2017} are able to reproduce our empirical IRX relation.  Such an analysis, however, goes beyond the scope of this work.

 In short, the universal relation of dust obscuration we found successfully reproduces IRX using  $Z$, $L_{\rm IR}$, $R_{\rm e}$ and $b/a$ of the local sample of SFGs with a minimal scatter of 0.17\,dex. More importantly, the distant SFGs up to $z=3$ are excellently unified by the same relation.

\section{SUMMARY} \label{sec:sec7}

Using a sample of $\sim$32\,000 local SFGs carefully selected from  SDSS, GALEX and WISE, we investigate the relationships between dust obscuration (i.e., IRX) and galaxy parameters, including stellar mass, SFR,  IR luminosity, the gas-phase metallicity, galaxy size and axial ratio, to pinpoint the key quantities regulating dust obscuration. We summarize our results as follows.

Our detailed analysis reveals that IRX correlates with SFR/IR luminosity, metallicity, galaxy size and axial ratio. In practice, the obscured SFR traced by IR luminosity yields the strongest correlation with IRX. This may partially be due to the fact that the IRX directly contains IR luminosity. We thus introduce an empirical relation for IRX in the form of a joint power-law function of $L_{\rm IR}$, $R_{\rm e}$ and $b/a$ with the power-low indices being a function of metallicity. The best fit to our SFG sample yields the empirical IRX relation described by Equation~\ref{eq:eq3} and \ref{eq:eq4}. The dispersion around this relation is surprisingly small, only about 0.17\,dex. The empirical IRX relation suggests that IRX generally increases with the $L_{\rm IR}$ and $Z$, and decreases with $R_{\rm e}$ and $b/a$. The power-law indices for $L_{\rm IR}$, $R_{\rm e}$ and $b/a$ are metallicity dependent, such that the sensitivity to these parameters reduces towards lower metallicity. The precise reasons for the flattening remain to be properly understood.

The SFGs in our sample are mostly disc-dominated galaxies ($n<2$). As a consequence, inclination effects bias the observed $L_{\rm UV}$ compared to what would be obtained from integrating over $4\pi$ steradians, inducing a weak dependence of the observed IRX on axial ratio.  We demonstrated that this inclination effect is a strong function of metallicity and can be quantitatively corrected using the empirical IRX relation. The genuine IRX can be obtained after removing the inclination-related bias (see Figure~\ref{fig:fig7}).

We show that the empirical IRX relation does not rely on galaxy stellar mass. Our results show that $Z$, $L_{\rm IR}$, $R_{\rm e}$ and $b/a$ are sufficient to reproduce IRX and no dependence on stellar mass is found in the residual between the observed and model IRX. Inclusion of a stellar mass term when fitting gives $IRX\propto M_\ast^{[-0.08, 0.17]}$ while other power-law indices remain unchanged within their respective uncertainties. 

Surprisingly, high-$z$ SFGs perfectly follow our empirical IRX relation in a population-averaged sense. No evolutionary effect is seen for SFGs out to $z$=3.  We point out that the key quantities regulating dust obscuration play the same role in both local and high-$z$ star-forming galaxies. The empirical IRX relation we obtained is indeed a universal relation of dust obscuration, independent from both stellar mass and cosmic epoch.

Our finding of the universal IRX relation reveals that IRX approximately increases with $L_{\rm IR}/R_{\rm e}^{[1.3 - 1.5]}$ instead of $L_{\rm IR}$ surface density. We argue that this might be attributed to the systematic difference between the spatial extent of the stellar distribution ($R_{\rm e}$) and that of star formation ($R_{\rm IR}$). If so, a homogeneous mixture model of dust and stars would be able to reproduce the universal IRX relation well in the high ($\sim$Solar) metallicity regime but would still fail at low metallicities. Alternatively, dust-star geometries more complex than a homogeneous mixture may need to be invoked. We stress that the gas-phase metallicity plays a fundamental role in shaping the IRX-related relationships.

\section*{acknowledgements}

We are grateful to the  anonymous referee for her/his useful comments and careful reading of the manuscript. This work is supported by the National Key Research and Development Program of China (2017YFA0402703), NSFC grant (11773076, 11703092), and the Chinese Academy of Sciences (CAS) through a grant to the CAS South America Center for Astronomy (CASSACA) in Santiago, Chile.  SW acknowledges support from the Chinese Academy of Sciences President's International Fellowship Initiative (grant no. 2017VMB001). ZP acknowledges the support from the Natural Science Foundation of Jiangsu Province (No.BK20161097).

\bibliographystyle{\mnras}

\bibliography{references}

\appendix

\section{Additional analysis results}\label{sec:appendix}

\begin{figure*}
\includegraphics[width=0.48\textwidth]{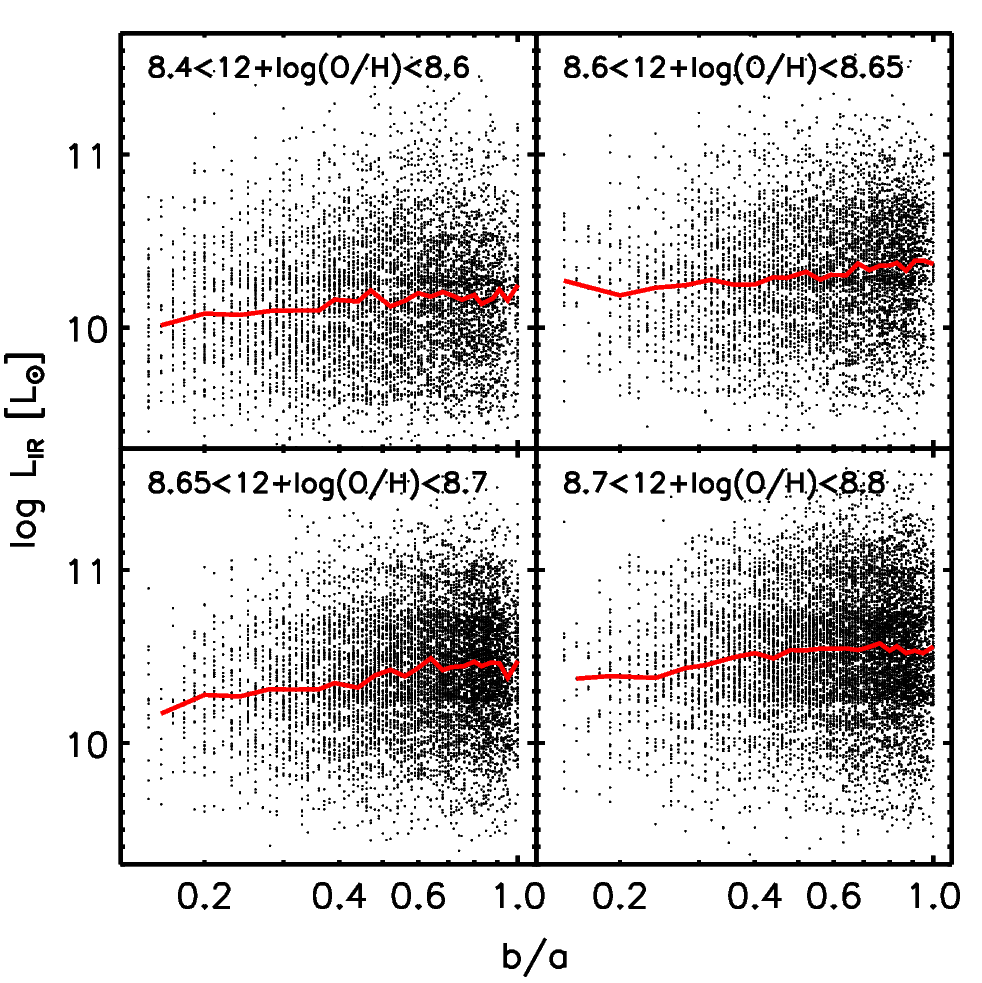}
\includegraphics[width=0.48\textwidth]{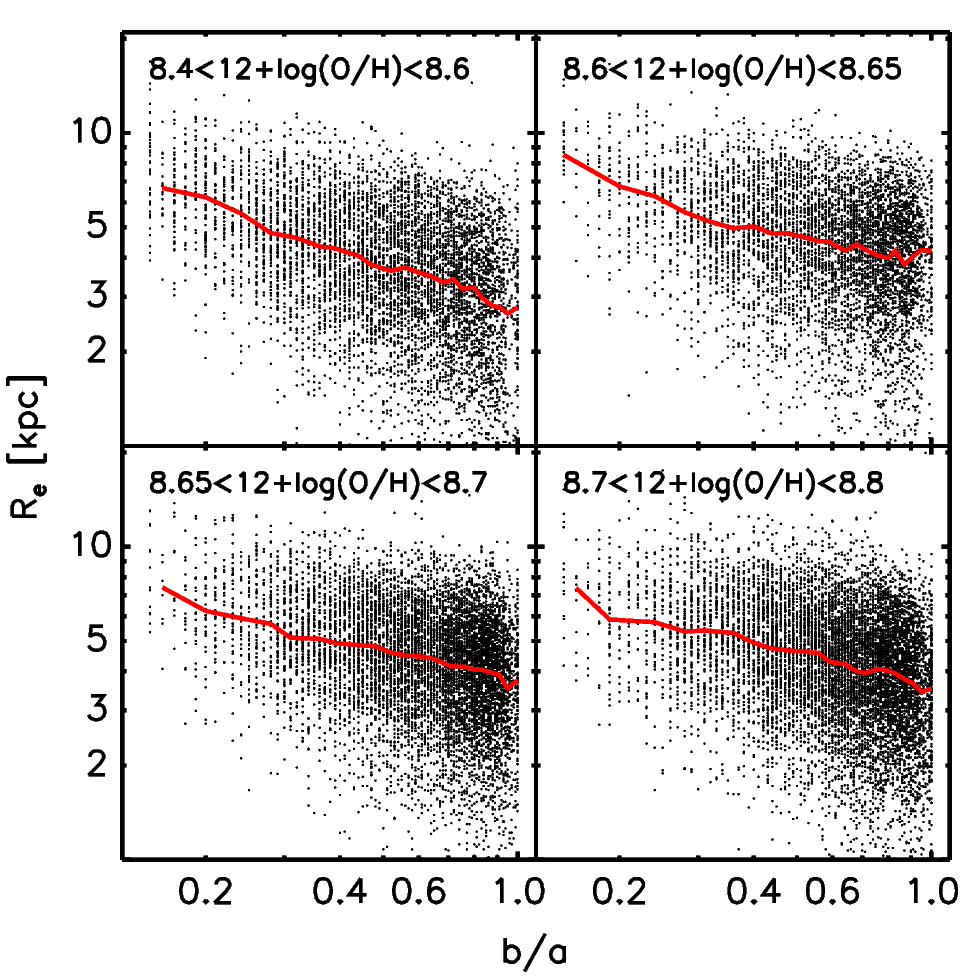}
\caption{IR luminosity $L_{\rm IR}$ (left plot) and the half-light radius $R_{\rm e}$ (right plot) as a function of axial ratio $b/a$ for our full sample of local SFGs, split into four metallicity bins. The red solid lines mark the median of data points. No dependence is found for $L_{\rm IR}$ on axial ratio. However, the SFGs of smaller sizes exhibit a lack of objects with low $b/a$, indicating that small SFGs tend to feature rounder shapes.}
\label{fig:figA1}
\end{figure*}

\begin{figure*}
\includegraphics[width=\textwidth]{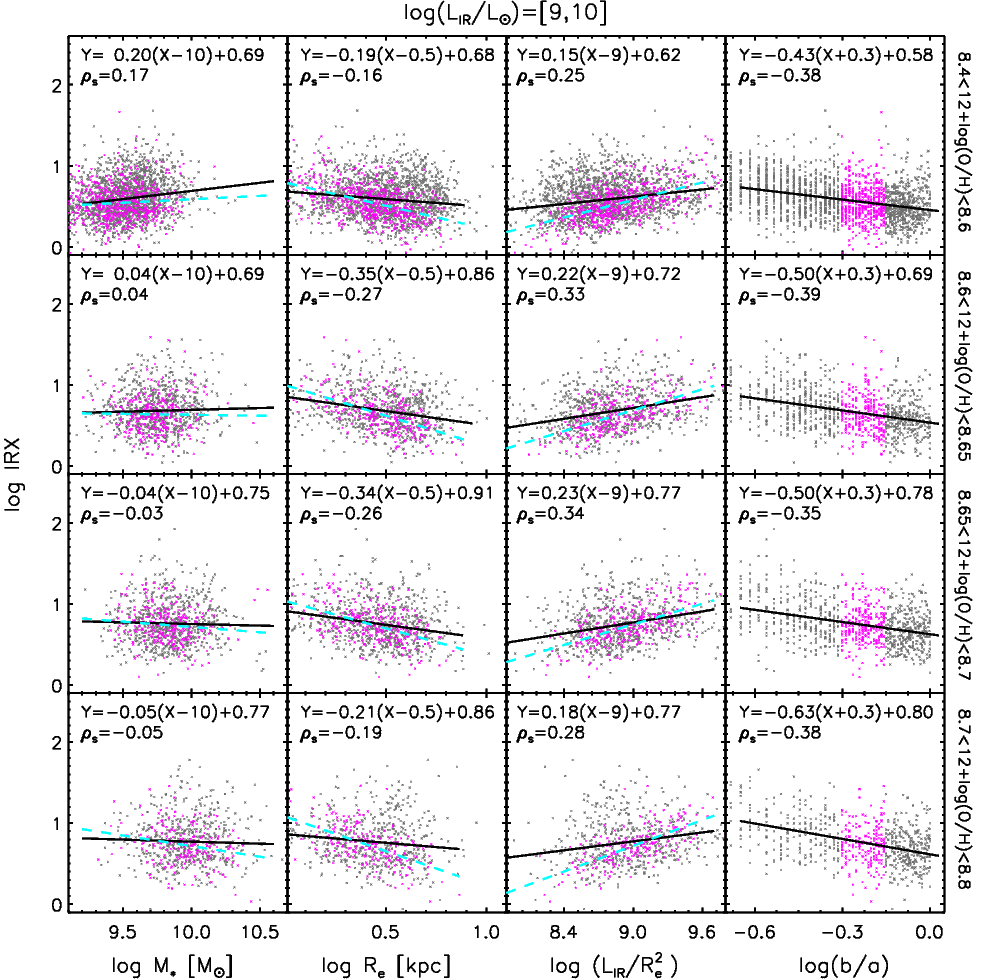}
\caption{Similar to Figure~\ref{fig:fig3} but for the local SFGs with 9$\le\log(L_{\rm IR}/L_\odot)<$10. }
\label{fig:figA2}
\end{figure*}

\begin{figure*}
\includegraphics[width=\textwidth]{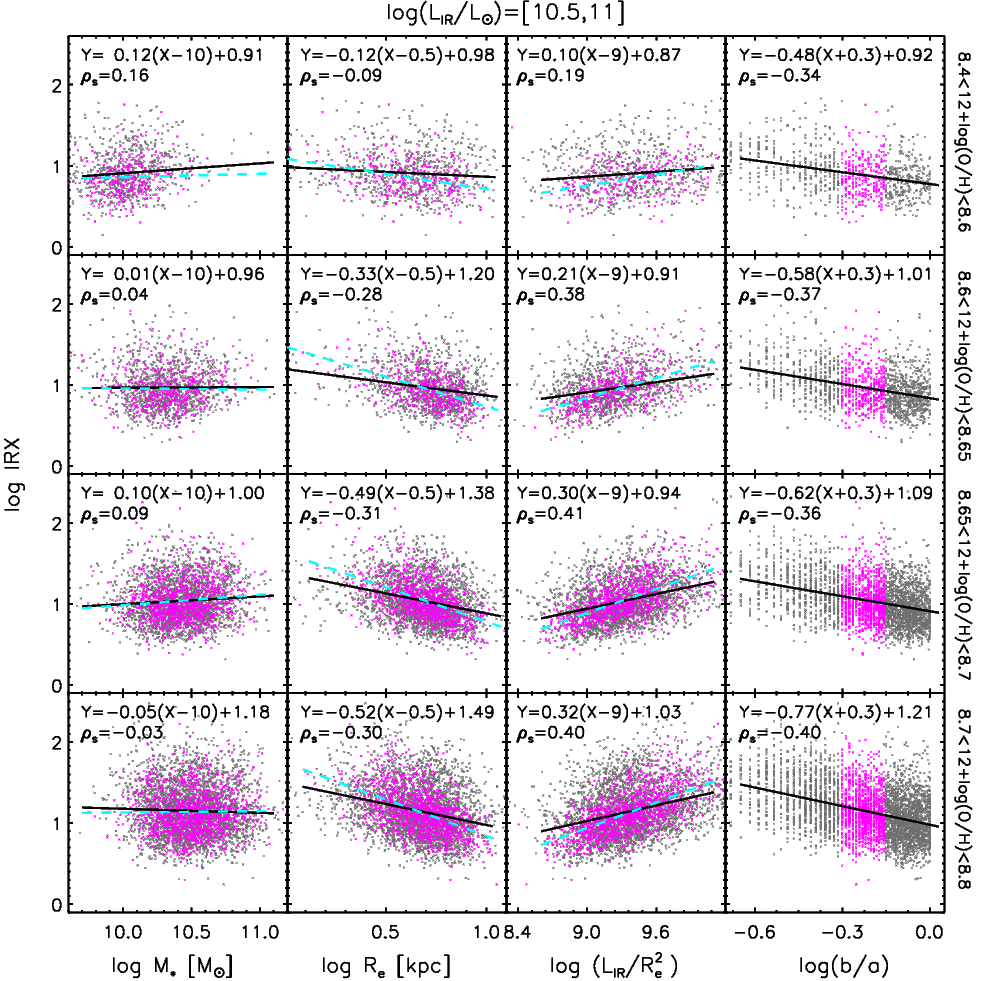}
\caption{Similar to Figure~\ref{fig:fig3} but for the local SFGs with 10.5$\le\log(L_{\rm IR}/L_\odot)<$11. }
\label{fig:figA3}
\end{figure*}

\begin{figure*}
\includegraphics[width=\textwidth]{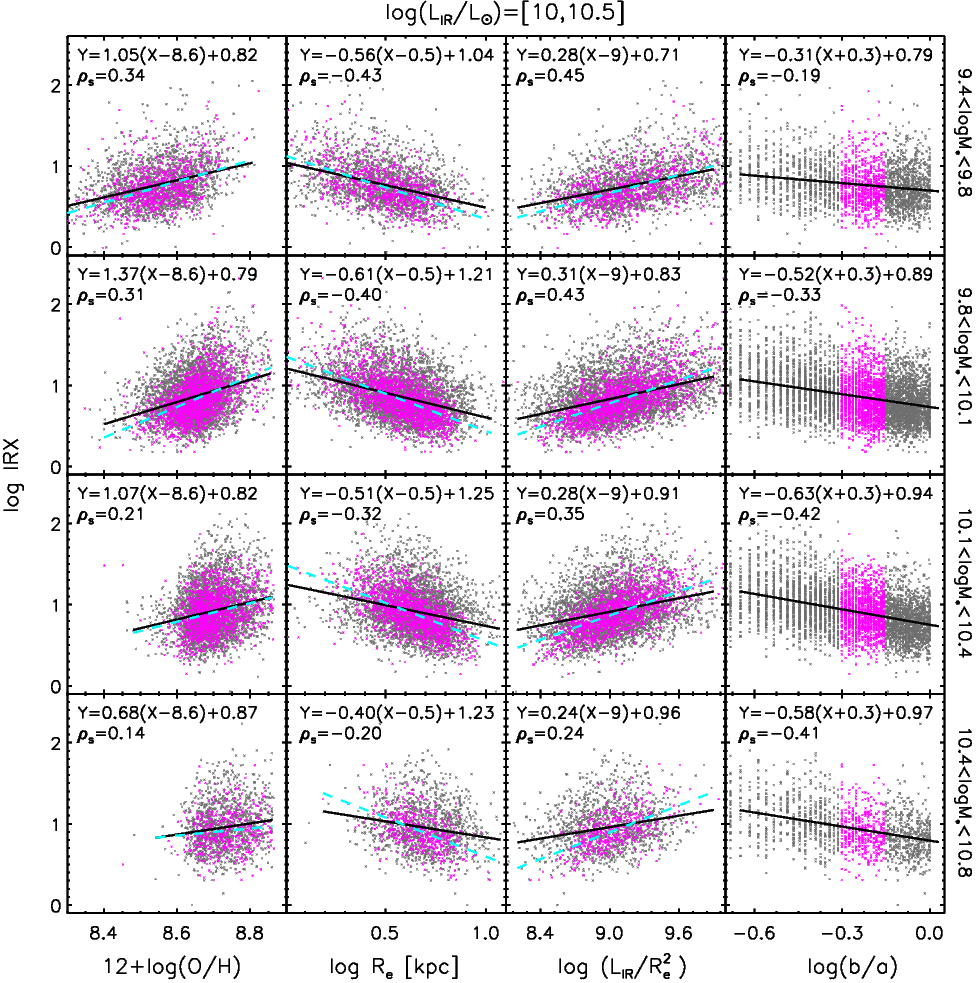}
\caption{Relationship between IRX and metallicity, size,  $L_{\rm IR}/R_{\rm e}^2$, axial ratio (from left to right) for a subsample of local SFGs with  10$\le\log(L_{\rm IR}/L_\odot)<$10.5 split into four stellar mass bins over 9.4$<$$\log M_\ast/M_\odot$$<$10.8 (from top to bottom). The meanings of symbols, color,  lines are the same as in Figure~\ref{fig:fig3}. At fixed IR luminosities and stellar masses, IRX still correlates with  $L_{\rm IR}/R_{\rm e}^2$, and anti-correlates with size and axial ratio, consistent with the same correlations at fixed IR luminosities and metallicities shown in Figure~\ref{fig:fig3}.  Apparently, metallicity becomes systematically higher at increasing stellar mass but the correlation between IRX and metallicity is strong with a steep slope when metallicity spans a sufficiently wide range as shown in the top-left two panels.}
\label{fig:figA4}
\end{figure*}

\begin{figure*}
\includegraphics[width=\textwidth]{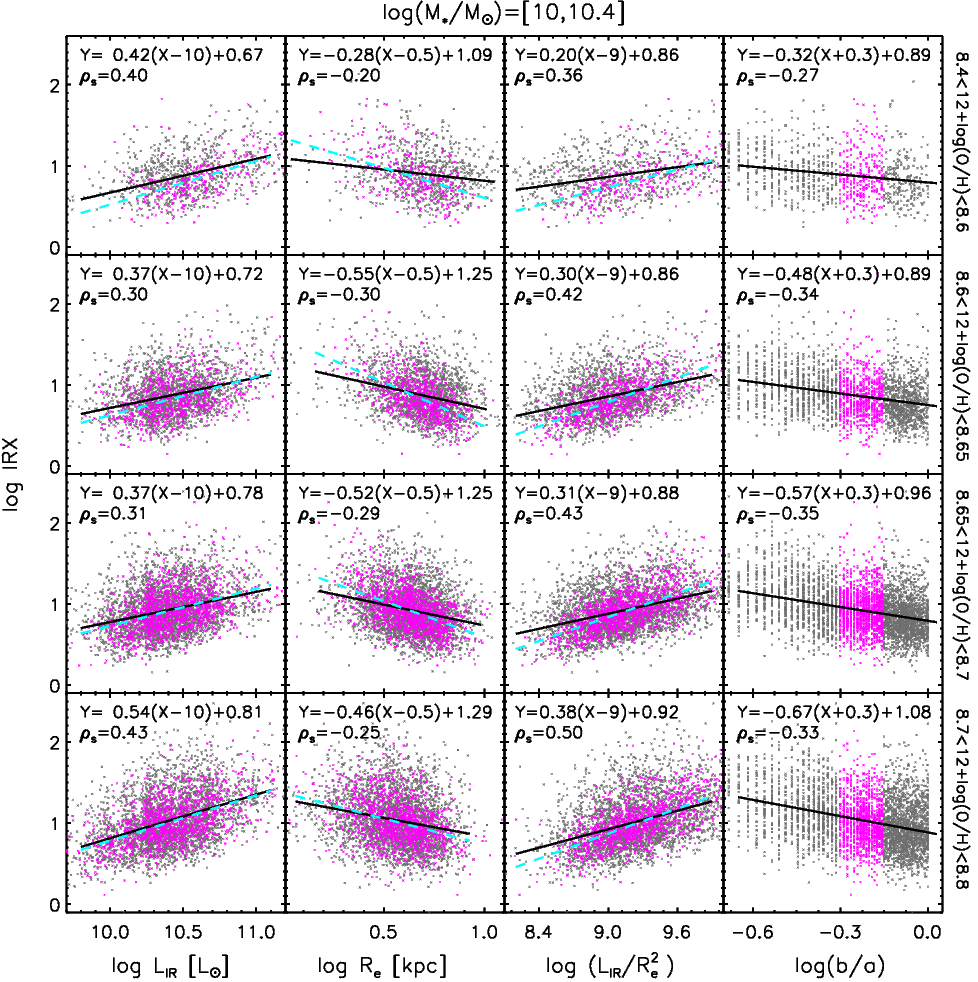}
\caption{Relationship between IRX and IR luminosity, size,  $L_{\rm IR}/R_{\rm e}^2$, axial ratio (from left to right) for a subsample of local SFGs with 10$<$$\log M_\ast/M_\odot$$<$10.4 split into four metallicity bins over 8.4$<$$\zoh$$<$8.8 (from top to bottom). The symbols, colors and line types follow those given in Figure~\ref{fig:fig3}.   IRX appears to correlate with both $R_{\rm e}$ and $L_{\rm IR}/R_{\rm e}^2$, and anti-correlate with $b/a$ at fixed stellar masses and metallicities, consistent with the same correlations presented in Figure~\ref{fig:fig3} and Figure~\ref{fig:figA4}.   It is clear that from the left panels that IRX tightly correlates with IR luminosity when controlling stellar mass and metallicity.}
\label{fig:figA5}
\end{figure*}

\begin{figure*}
\includegraphics[width=\textwidth]{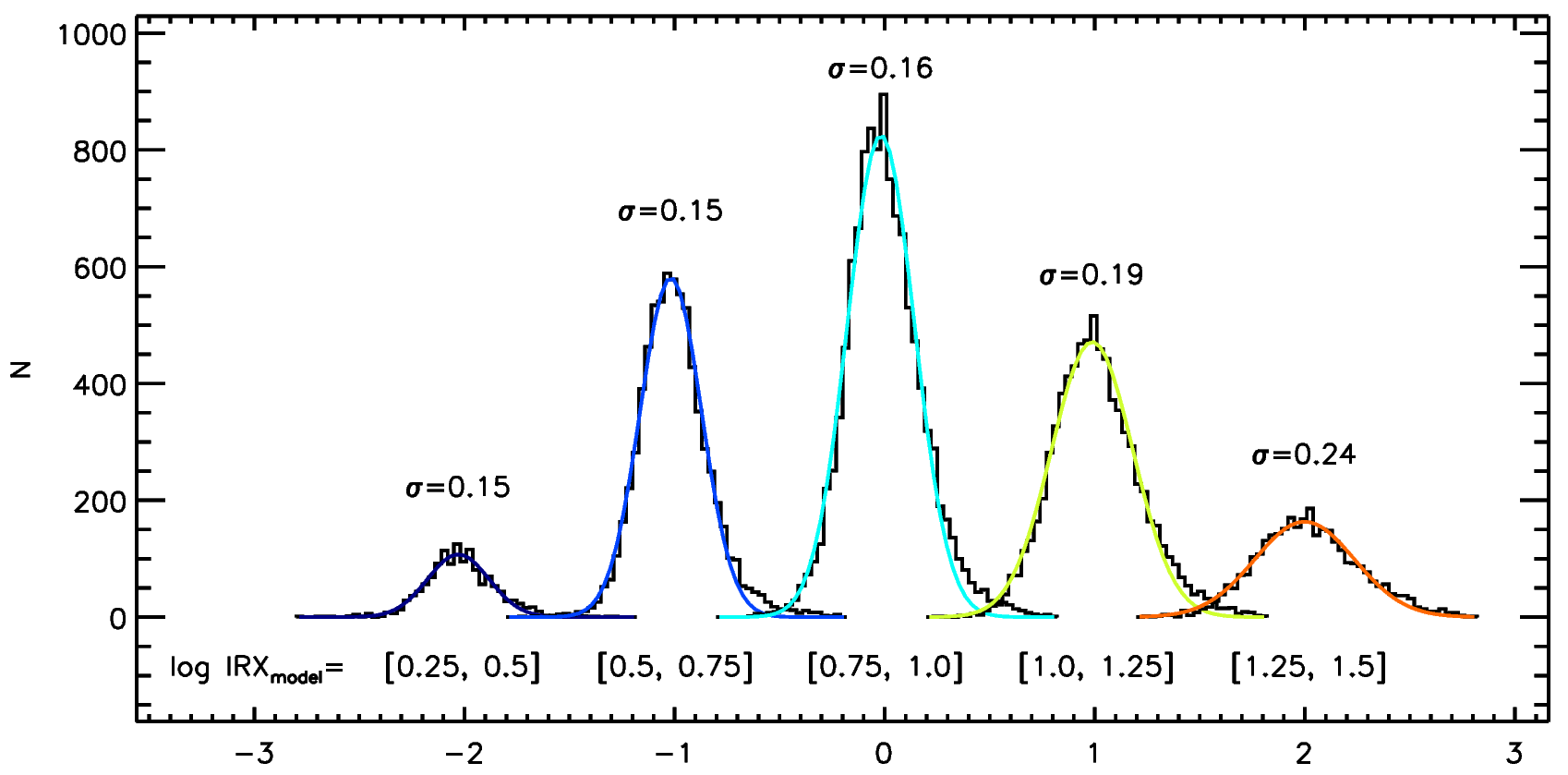}
\caption{Distributions of IRX residuals for SFGs in five $\log IRX_{\rm model}$ bins from $\log IRX_{\rm model}$= 0.25 to 1.5. The five distributions are shifted by -2, -1, 0, +1, +2 respectively for clarity. The best-fitting Gaussian functions are also shown. The larger $IRX_{\rm model}$ bins have higher dispersion, with $\sigma=$ 0.15 to 0.24\,dex.}
\label{fig:figA6}
\end{figure*}

\begin{figure*}
\includegraphics[width=\textwidth]{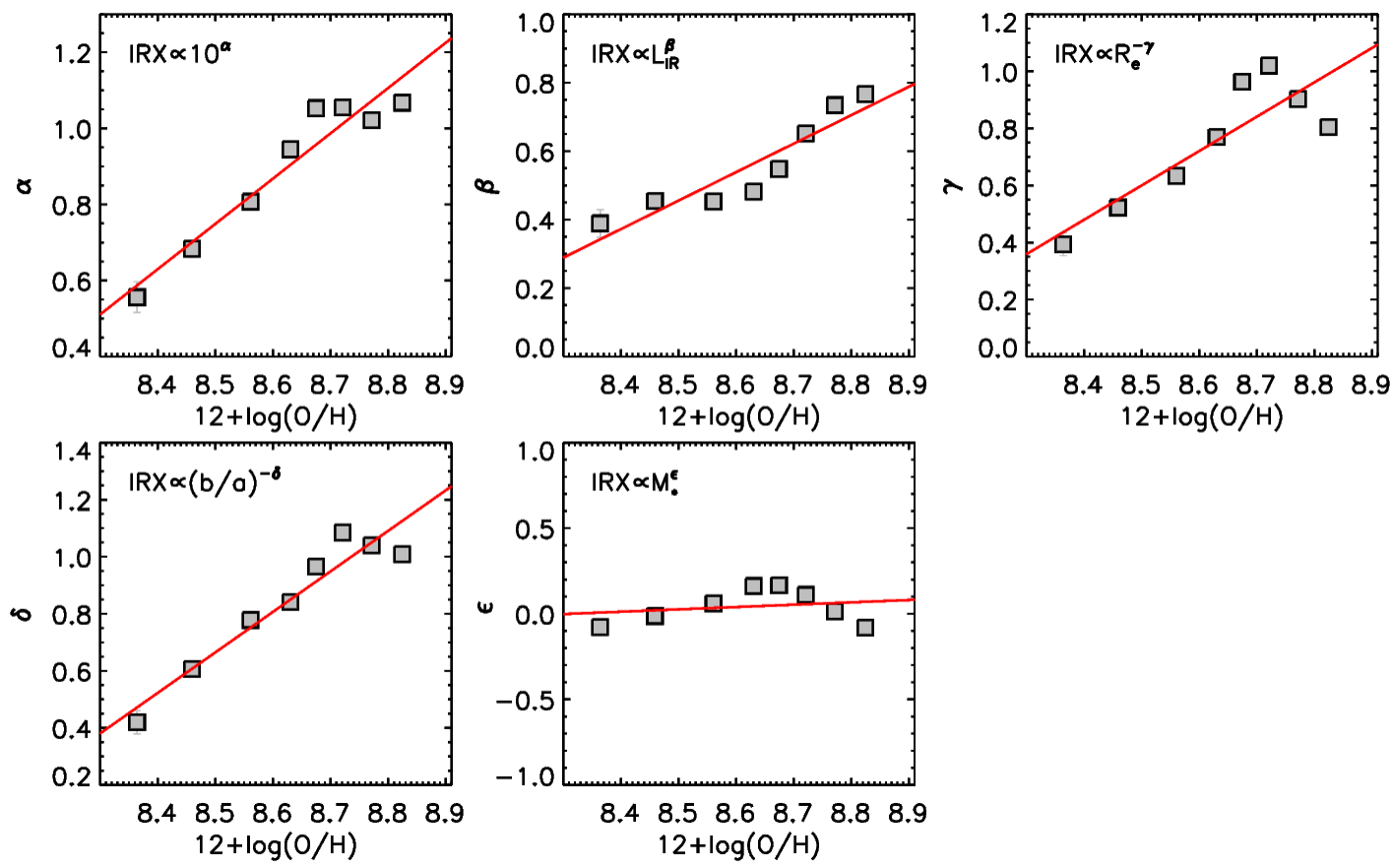}
\caption{The best-fitting power-law indices $\alpha$, $\beta$, $\gamma$, $\delta$ and $\epsilon$ given in Equation~\ref{eq:eq6} as a function of metallicity. Their best-fitting relations are shown with red solid lines. }
\label{fig:figA7}
\end{figure*}

Figure~\ref{fig:figA1} to Figure~\ref{fig:figA7} present additional  results from our analysis to support our conclusions in the text.  See the captions of each figure for more details. 

\bsp	
\label{lastpage}
\end{document}